\documentclass[aps,twocolumn,pra,showpacs,superscriptaddress]{revtex4-1}
\usepackage{graphicx,color}
\usepackage{amsmath}
\usepackage{amssymb}
\usepackage{epstopdf}
\usepackage{hyperref}
\usepackage{color}

\usepackage{braket}
\usepackage{mathtools}
\newcommand{\ie}{\textit{i}.\textit{e}.~}

\begin{document}
\title{Efficient experimental design of high-fidelity three-qubit
quantum gates via genetic programming}
\author{Amit Devra}
\affiliation{Department of Physical Sciences, Indian
Institute of Science Education \& 
Research (IISER) Mohali, Sector 81 SAS Nagar, 
Punjab 140306 India.}
\author{Prithviraj Prabhu}
\affiliation{Sri Sivasubramaniya Nadar College of Engineering, 
Kalavakkam, Chennai, Tamil Nadu 603110
India.}
\author{Harpreet Singh}
\affiliation{Department of Physical Sciences, Indian
Institute of Science Education \& 
Research (IISER) Mohali, Sector 81 SAS Nagar, 
Punjab 140306 India.}
\author{Arvind}
\affiliation{Department of Physical Sciences, Indian
Institute of Science Education \& 
Research (IISER) Mohali, Sector 81 SAS Nagar, 
Punjab 140306 India.}
\author{Kavita Dorai}
\affiliation{Department of Physical Sciences, Indian
Institute of Science Education \& 
Research (IISER) Mohali, Sector 81 SAS Nagar, 
Punjab 140306 India.}
\email{kavita@iisermohali.ac.in}
\begin{abstract}
We have designed efficient quantum circuits for the
three-qubit Toffoli (controlled-controlled NOT) and the
Fredkin (controlled-SWAP) gate, optimized via genetic
programming methods. The gates thus obtained  were
experimentally implemented  on a three-qubit NMR quantum
information processor, with a high fidelity.  Toffoli and
Fredkin gates in conjunction with the single-qubit Hadamard
gates form a universal gate set for quantum computing, and
are an essential component of several quantum algorithms.
Genetic algorithms are stochastic search algorithms based on
the logic of natural selection and biological genetics and
have been widely used for quantum information processing
applications. The numerically optimized rf pulse profiles of
the three-qubit quantum gates achieve $> 99\%$ fidelity.
The optimization was performed under the constraint that the
experimentally implemented pulses are of short duration and
can be implemented with  high fidelity. Therefore the gate
implementations   do not suffer from the drawbacks of rf
offset errors or debilitating effects of decoherence during
gate action. We demonstrate the advantage of our pulse
sequences by
comparing our results with existing 
experimental schemes.
\end{abstract}
\maketitle
\section{Introduction}
\label{intro}
Quantum technologies that have been proposed to build
quantum computers should be able to achieve a high degree of
control over a universal set of quantum gates that form the
basic elements of quantum circuits~\cite{nielsen-book-02}.
Any quantum computing circuit can be realized using a
universal set of two-qubit gates and a set of local
unitaries~\cite{barenco-pra}.  However using this basic set
of single- and two-qubit gates to decompose multiqubit
unitary propagators for large qubit registers, leads to
problems of scalability and decoherence due to long
operation times of the circuits~\cite{sleator-prl-95}.
Hence, the idea of multi-level quantum logic was developed
which used three- and four-qubit quantum gates to
considerably simplify the quantum
circuit~\cite{mikko-prl-04,mohammadi-qip-08}.

Three-qubit gates such as the Toffoli gate (which is
equivalent to controlled-controlled-NOT operation) and the
Fredkin gate (which is equivalent to a controlled-SWAP
operation) play an important role in quantum
circuits~\cite{smolin-pra-96},
fingerprinting~\cite{buhrman}, optimal
cloning~\cite{hofmann} and quantum error
correction~\cite{cory-qec,nature-qec}.  The Fredkin gate was
discussed early on as a useful gate for optical
implementations of quantum
computing~\cite{milburn-prl-89,shamir-86}.  The Toffoli and
the Fredkin gates, in conjunction with the single-qubit
Hadamard gate, form a universal set of quantum
gates~\cite{peres-pra-85,shi-03}.  Previous implementations
of these universal three-qubit gates relied on their
decomposition into sets of single- and two-qubit
gates~\cite{chau-prl-95,shende-09}.  Efficient construction
of three- and four-qubit gates using an optimal set of
global entangling gates has been recently
explored~\cite{ivanov-pra-15}, and a machine learning type
of algorithm has been used to design high-fidelity
single-shot three-qubit gates which do not require prior
decomposition into sets of two-qubit
gates~\cite{zahedinejad-prl-15,sanders-16}.

Three- and four-qubit gates were experimentally realized
early on in NMR quantum computing by several
groups~\cite{fredkin-chinese,du-pra-01,das-jmr-02,zhang-joptb-05}.
The Toffoli gate has been experimentally implemented using
trapped ions~\cite{monz-prl-09} and circuit
QED~\cite{qed-prb-12} and superconducting
qubits~\cite{toffoli-supercon,toffoli-superconducting}.
Several studies of the Toffoli gate have focused on its
experimental realization using optical
setups~\cite{lanyon-nature,optical-toffoli-prl,optical-toffoli-pra}.
Other implementations of the Toffoli gate include an optimal
version using a reduced set of two-qubit
gates~\cite{optimized-toffoli,toffoli-scirep-15}.  The
Fredkin (controlled-SWAP) gate was recently experimentally
realized using photonic qubits~\cite{ctrlswap-scirep-17}.

Several optimization techniques have been successfully
developed for quantum control such as strongly modulated
pulses~\cite{fortunato-jcp-02}, GRAPE
optimization~\cite{tosner-jmr-09,glaser-review}, sequential
convex programming~\cite{qtmgates-13} and optimal dynamical
discrimination~\cite{long-homonuclear}.  A novel set of
optimization techniques broadly categorized as `Genetic
Algorithms (GAs)', have also been proposed as a means to
achieve a global minimum for the
optimization~\cite{holland-book}.  GAs borrow their
optimization protocol from the basic tenets of evolutionary biology,
wherein the breeding strategy of a population is to increase
the fitness levels and offspring-producing capability of
individuals by crossing over of genetic
information~\cite{forrest}.  In quantum information
processing, GAs have been used to optimize quantum
algorithms~\cite{suter-ia-08,hardy-2010,bang-korean} and
quantum entanglement~\cite{navarro-pra-06}, for optimal
dynamical decoupling~\cite{lidar-pra}, and to optimize
unitary transformations for a general quantum
computation~\cite{manu-pra-12,manu-pra-14}.

In this work, we explore the efficacy of GAs in optimizing
the Toffoli and Fredkin gates alongwith a set of
single-qubit gates, on a three-qubit NMR quantum information
processor.  We design an implementation of these gates which
uses only ``hard'' (\ie short duration) rf pulses of
arbitrary flip angles and phases, punctuated by intervals of
evolution under the system Hamiltonian.  We are hence able
to substantially avoid the pitfalls associated with ``soft''
shaped NMR pulses, namely pulse calibration errors and
degradation due to decoherence occurring during the long
gate times of such pulses. 
The constraints are put in from practical considerations,
whereby we want to design the gate using only 
a certain kind of short duration rf pulses,
and then genetic algorithms are used to optimize the
protocol.
We compared our experimental
results with previous NMR implementations of these
three-qubit gates using standard transition-selective shaped
pulses.  We demonstrate that our scheme is substantially
better, with obtained experimental fidelities $\approx 14$\%
higher as compared to the standard implementations and also
report a 5-6 times savings in gate implementation time as
compared to the standard implementations.  Our scheme is
general and can be implemented on any quantum hardware to
generate circuits for three-qubit gates of high fidelity.

The material in this paper is arranged as follows:
In Section~\ref{theory} we describe the optimization
scheme based on GAs. In Section~\ref{exptmental} we discuss
the implementation of optimized gates on an NMR system of
three qubits with Section~\ref{system} containing the details of
the NMR system, Section~\ref{ninety} describing the optimized
implementation of a $90^\circ$ pulse, Section~\ref{cnot} the
implementation of a two-qubit CNOT gate. While Sections~\ref{fredkin}
and~\ref{toffoli} describe the implementation of the optimized
Fredkin and Toffoli gates respectively, in  Section~\ref{compare} we
compare our results with standard implementations of
these gates. 
Section~\ref{concl} offers some concluding remarks.
\section{Numerical Optimization of Three-Qubit Gates via
Genetic Programming} \label{theory} Unitary operators
corresponding to controlled operations and to quantum gates
can be implemented on an NMR quantum information processor
by a suitable set of radiofrequency pulses of a specific
frequency, amplitude and phase, interspersed with delays
which correspond to free evolution under the system
Hamiltonian.  The problem of numerical optimization of any
quantum gate can hence be recast as an optimization problem
in genetic programming, wherein the fitness function to be
optimized depends on the target unitary operator, with its
corresponding set of pulse parameters and delay times.  The
fitness function which determines the relative distance
between two operators, is defined in our scenario
as~\cite{souza-prl-11,manu-pra-14}: \begin{equation} {\cal
F} = \frac{\vert {\rm Tr}(U_{{\rm tgt}} U^{\dagger}_{{\rm
opt}}) \vert} {\sqrt{ {\rm Tr}(U_{{\rm tgt}}
U^{\dagger}_{{\rm tgt}}) {\rm Tr}(U_{{\rm opt}}
U^{\dagger}_{{\rm opt}}) }} \end{equation} where $U_{{\rm
tgt}}$ is the target unitary operator of the desired gate to
be optimized and $U_{{\rm opt}}$ is the actual operator
generated by the GA optimization.  The fitness function is
normalized such that when $U_{{\rm opt}}$=$U_{{\rm tgt}}$,
the fitness has the maximum value of unity.

The derived unitary operator of the
gate to be optimized, $U_{\rm opt}$, is defined as:
\begin{eqnarray}
U_{\rm opt} &=& \prod_{l=1}^{N} \exp[-i({\cal
H}_\mathrm{NMR}+\Omega I_{\phi_{l} k})\tau_{l}] \exp[-i{\cal
H}_\mathrm{NMR}\delta_{l}] \nonumber \\
I_{\phi_{l}} &=& \dfrac{1}{2}\left(\sigma_x 
\cos\phi_{l} +\sigma_y \sin\phi_{l} \right)
\label{target}
\end{eqnarray} 
where $\Omega$ denotes the amplitude of the
rf pulse, $\phi_l$ is the phase of the $l$th
rf pulse, $\tau_l$ is the pulse length,
$\delta_l$ denotes an evolution period under the
system Hamiltonian, and
$\sigma_x$ and $\sigma_y$ are the Pauli
$x$ and $y$ matrices respectively.
The first term in the expression for the
desired unitary operator $U_{{\rm opt}}$ (Eqn.~\ref{target}) 
describes the system and RF Hamiltonians,
while the second term 
describes the evolution under the  
the system Hamiltonian. 
The system Hamiltonian ${\cal H}_\mathrm{NMR}$ in
the rotating frame is given by
\begin{equation}
{\cal
H}_\mathrm{NMR}=-\pi\sum_{i=1}^n(\nu_i-\nu_{rf}^{i})\sigma^i_z
+ \sum_{i<j,=1}^n \frac{\pi}{2} J_{ij}\sigma^i_z\sigma^j_z
\end{equation}
where $n$ denotes the number of spins, $\nu_i$ and
$\nu_{rf}^{i}$ are the chemical shift and the rotating frame
frequencies respectively, $J_{ij}$ are the scalar coupling
constants and $\sigma_{z}$ is the Pauli $z$ matrix.

\begin{figure}[h]
\centering
\includegraphics[angle=0,scale=1.0]{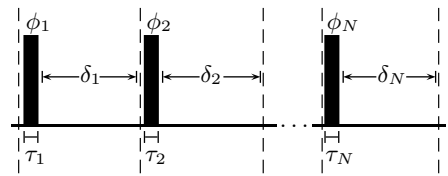}
\caption{Desired unitary propagator 
is represented by a set of $N$ pulses of pulse width
$\tau_l$ and phase $\phi_l$, punctuated by
$N$ delays of interval
$\delta_l$ ($l= 1...N$).}
\label{propagator}
\end{figure}

We choose to decompose the desired unitary operator
$U_{{\rm opt}}$ as a set of $N$ hard (\ie high-power,
very short duration) rf pulses, each of fixed amplitude
$\Omega$, pulse length $\tau_{l}$ and phase $\phi_{l}$, and
a set of $N$ delays, each of interval
$\delta_{l}$ in duration.  This set of pulses and
delays denotes the basic propagator (Fig.~\ref{propagator}).   
These optimal pulse phase, pulse width and delay
values (which together
constitute one possible solution to
the optimization problem) 
can be encoded in the form of a matrix
of order
$N \times 4$ where the number of rows ($N$) specifies the number
of operations that the desired unitary operator is
decomposed into.
By increasing the number of rows, we
increase the accuracy and control.
The values of the four columns of the matrix are
detailed below:
\begin{itemize}
\item{\bf Column 1:} 
represents the width ($\tau$) of a hard pulse, which
is used to tune the value of the angle of rotation of
the pulse ($\theta \in \{0,2\pi\}$). 
\item{\bf Column 2 and 3:} 
The phase of rotation
of the rf pulse is represented in these two columns.
The second column takes the values 
either 0 or 1, implying a positive
or negative phase ($\phi \in \{0,\pi\}$ or
$\phi \in \{-\pi,0\}$, respectively). 
The third column contains the entire
range of $\phi$ values from
$\{0,2\pi\}$.
\item{\bf Column 4:} 
The values in this column represent the time evolution
$\delta_l$ between hard pulses. The maximum value that an
element in this column can assume is relative to the type of
gate chosen.  A Fredkin gate inherently requires more time
than a CNOT gate, and hence will be given more freedom in
choosing delay lengths.
\end{itemize}
Due to accuracy constraints imposed by the NMR hardware on
which the gates are implemented, we can only obtain a
resolution of 0.01 degrees for the phase, and a resolution
of 1$\mu$s for the delay. Hence the values in our results are
discretized accordingly. It should also be noted that
the power of each hard rf pulse is fixed and is not optimized.

As the first step in the genetic algorithm, an initial
population of solutions, \ie $n$ `randomly chosen
chromosomes' is created. Considering we run the algorithm
using a variable number of rows, we must first decide upon a
suitable population size to run the algorithm with. As the
number of rows increases, so does the time taken to convert
a matrix of rf pulses to a gate matrix. We thus used
population sizes ranging from 350 to 750, for rows ranging
from 3-20.  There are three main operations which form the
backbone of the genetic algorithm as
described below~\cite{holland-book}:

\begin{itemize}
\item {\bf Selection:} Selecting individuals for
crossover and mutation processes is important as
it dictates the direction taken by the population in the
fitness landscape~\cite{holland-book}.  We initially 
use a low selection pressure in order to explore
all possible candidate solutions.
If a viable solution is recognized,
the intensity of the selection pressure is increased, 
to allow for exploitation of neighbors of the
recognized solution.  After attempting existing selection
mechanisms such as roulette, rank, tournament and stochastic
acceptance~\cite{jebari-selection}, we devised our 
own selection mechanism which we
call ``Luck-Choose''. 
The operation involves
first multiplying pseudo-randomly generated weights to the
fitness values of all individuals, and subsequently
determining the highest among the output values.
Using the Luck-Choose method, the algorithm converged to a
solution much faster.
\item{\bf Crossover:} 
The crossover
operation in the genetic algorithm method swaps congruent parts of
individual members of the population as follows: Two members are
chosen from the population using the Luck-Choose 
selection method.  Two
numbers are randomly chosen within the maximum number of
rows, and two numbers are randomly chosen within the maximum
number of columns.  The first number of each corresponds to
the starting point of the crossover and the second number
corresponds to the end point.  Using the above four numbers
we create a rectangular sub-matrix, which is swapped between
both the selected individuals.  In addition, we added
another operation called {\em flip} in order to address the
problem of non-commutativity of rf pulses. The {\em flip}
operation selects a single member using the Luck-Choose
method and swaps its constituent rows.
\item{\bf Mutation:} This operation depends
heavily on the amount of stochastic noise required.
Stochastic noise adds a random amount of noise to
ensure that the algorithm does not stagnate at any of the
local optima.  In the initial stages, low stochastic noise
is preferred, so the mutation operation may be disabled. However
after the algorithm explores
the population landscape through a few
generations, the chances of getting stuck in local optima
increase. The probability of mutations is
then increased in steps upto a threshold
value, above which the stochastic noise would only serve to
drive candidate solutions away from the global optimum.
Mutation takes a single member, selected using the
Luck-Choose method, and changes all its data values.
\end{itemize}
After running the genetic algorithm, outputs are obtained
with fidelities in the lower 0.80 range.  In order to
increase the fidelity, we used the concept of a localized
optimizer, which is a GA tool that optimizes only within a
very small region of the fitness landscape. This is done by
localizing the range of values that constituent chromosomes
can take. As the maximum fidelity increases, we increase the
selection pressure to further minimize the region of
optimization of the algorithm, in the fitness landscape. The
chromosomes from the main optimizer, which yield fidelity
greater than 0.8 are then passed through this local
optimizer to increase the fidelity. In the general case, we
let the local optimizer run for 1000 seconds. If the
fidelity crossed 0.99, the solution was deemed acceptable.
In certain cases, local optimizer runtimes were further
increased, to increase the final fidelity.
Table~\ref{tab1a} gives details of the runtime per iteration
in the main and local optimizers, as well as the
corresponding count of Floating Point Operations per Second
(FLOPS), for the optimization of a $90^{\circ}$
spin-selective rf pulse.  All the rows for which the pulse
duration is mentioned in the table have a fidelity greater
than 0.99.  The genetic algorithm was performed using
MATLAB~\cite{matlab}. An iteration of the program running
the algorithm for 15 rows and 500 chromosomes, took an
average time of 3 hours using a single core for processing,
on an i7-4700MQ processor with 8 GB of RAM. For parallel
processing, the Parallel Computing Toolbox was used,
enabling us to run 6 iterations simultaneously on 6 virtual
cores for approximately 4 hours. This reduced the average
runtime per iteration to approximately 40 minutes.  The
local optimizer however was run from 10 minutes to 15 hours
depending on the final fidelity required and the fidelity of
the starting matrix.
\begin{table}[t]
\centering
\begin{tabular}{|p{0.7cm}|p{2.6cm}|p{2.6cm}|p{1.3cm}| }
\hline
Row &
Main Optimizer&Local Optimizer &
Pulse \\
No.&
Time(s)/GFLOPS&Time(s)/GFLOPS& Duration\\
\hline  
~~1& ~277.8/901.5 & ~NA          &~NA\\  
~~2& ~559.1/1814.2 &~NA          &~NA\\
~~3& ~871.3/2827.3  &~978/3173.6 &~101.4$\mu$s \\
~~4& ~1183.8/3841.4 &~46.16/149.8 &~146.8$\mu$s  \\
~~5& ~1435.8/4656.5 &~59.66/193.6  &~164.8$\mu$s  \\
~~6& ~1751.5/5683.6 &~28.5/92.5  &~253.7$\mu$s \\
~~7& ~2005.6/6508.1 &~23.2/75.3  &~243.7$\mu$s \\
~~8& ~2176.6/7063.1 &~19.2/62.3  &~292.3$\mu$s \\
\hline
\end{tabular}
\caption{Table of total optimization
time and total pulse duration against the number of rows for 
the optimization of a $90^{\circ}$
spin-selective rf pulse. The optimization time
per iteration is shown, alongwith the corresponding 
number of FLOPs used.}
\label{tab1a}
\end{table}
\section{Experimental Implementation of Numerically Optimized Gates}
\label{exptmental}
\subsection{Experimental NMR qubits}
\label{system}
\begin{figure}[h]
\centering
\includegraphics[angle=0,scale=1.0]{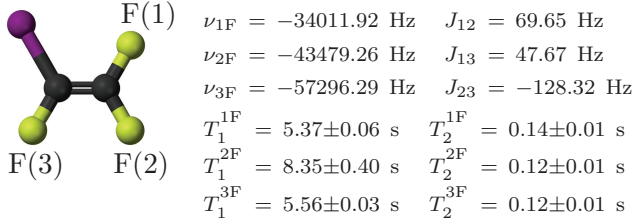}
\caption{Structure of iodotrifluoroethylene molecule with
measured values of chemical shifts ($\nu_{i}$) and scalar
couplings (J).}
\label{molfig}
\end{figure}

The three fluorine ($^{19}\text{F}$) spins of the molecule
iodotrifluoroethylene were used to encode the three 
qubits (Fig.~\ref{molfig}).
The three qubits were initialized into a
pseudopure state $\Ket{110}$ via the spatial averaging
technique~\cite{cory-physicad} with the density operator given by
\begin{equation}
 \rho_{110} = \dfrac{1-\epsilon}{8}I + \epsilon \Ket{110} \Bra{110}
\end{equation}
where thermal polarization ($\epsilon$) is approximately
$10^{-5}$ and $I$ is a $8 \times 8$ identity matrix. The
experimentally created pseudopure state was tomographed with
a fidelity of 0.97. All the experimental density matrices
were reconstructed using a reduced tomographic
protocol~\cite{leskowitz} and
maximum likelihood estimation~\cite{singh-pla-16} with the set of operations
given by $\{III, IIY, IYY, YII, XYX, XXY, XXX\}$,
where $I$ is the identity operation, $X$ and $Y$ are the
single spin angular momentum operators which can be implemented 
by applying a $\pi/2$ pulse on the corresponding spin.  
The operators for
tomographic protocols were numerically optimized using
genetic programming, each having a length of approximately
200 $\mu s$ and an average fidelity of $\geq$ 0.99. The
fidelity of the experimental density matrix was computed by
measuring the projection between the theoretically expected and
experimentally measured states using the Uhlmann-Jozsa
fidelity measure~\cite{uhlmann,jozsa}
\begin{equation}
  F = {\rm Tr} \left( \sqrt{\sqrt{\rho_{{\rm th}}}\rho_{{\rm exp}}
\sqrt{\rho_{{\rm th}}}} \right)
\end{equation}
where $\rho_{{\rm th}}$ and $\rho_{{\rm exp}}$ denote the
theoretical and experimental density matrices
respectively. 

Since we used a system of three homonuclear (same spin
species) spins, the control on all three spins happens
simultaneously and the optimization operator
$U_{\rm opt}$ is modified as:
\begin{equation}
\begin{split}
U_{\rm opt} =& \prod_{l=1}^{N} \exp[-i({\cal H}_\mathrm{NMR}+\Omega (I_{\phi_{l} 1}+I_{\phi_{l} 2}+I_{\phi_{l} 3}))\tau_{l}]\\
           & \exp[-i{\cal H}_\mathrm{NMR}\delta_{l}]
\end{split}
\end{equation}
The amplitude ($\Omega$) of
the hard rf pulse was kept fixed at$120.88\times10^{3}$ rad/s, the
hard pulse flip angle was taken in range of $\{0,3\pi/2\}$ and
range for length of the pulse ($\tau$) was adjusted
according to these two factors. The value of the delay between the
pulses was chosen depending upon the
unitary being optimized.
\subsection{Implementation of $90^{\circ}$ selective rf pulse}
\label{ninety}
\begin{figure}[h]
\centering
\includegraphics[angle=0,scale=1.0]{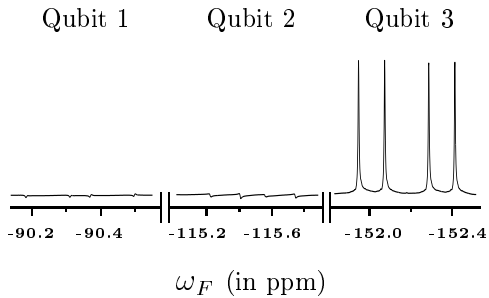}
\caption{Experimental implementation of
a numerically optimized $90^{\circ}$ selective pulse on 
the third qubit along the $Y$-axis, applied on a
thermal equilibrium state.}
\label{fig3}
\end{figure}

\begin{table}[h]
\begin{tabular}{|c|c|c|c| }
\hline
$\mathbf{l}$&$\mathbf{\tau_l(\mu s)}$&$\mathbf{\phi_l}$&$\mathbf{\delta_l(\mu s)}$ \\
\hline  
~1~~& ~16 & ~87.65~ &21 \\

~2~~&~33  &~269.97~ &21\\

~3~~&~16  &~92.29~  &0\\
\hline
\end{tabular}
\caption{Table representing the pulse sequence for selective
pulse. First column represents the number of propagators.
The second,
third and fourth columns give the pulse width ($\tau$),
phase ($\phi$) and delay ($\delta$) values, respectively.}
\label{taba}
\end{table}
To rotate a single spin in a homonuclear system
we need
a selective excitation pulse. We optimized the pulse
sequence via genetic algorithm for a $90^{\circ}$ selective pulse
on the third qubit along the $Y$-axis, using only  
hard pulses and delays.  The unitary for the selective
pulse is given by, 
\begin{equation}
    U_{\rm tgt}= \dfrac{1}{\sqrt{2}}
 \begin{bmatrix}
    1 & -1 & 0 & 0 & 0 & 0 & 0 & 0 \\
    1 &  1 & 0 & 0 & 0 & 0 & 0 & 0 \\
    0 & 0 & 1 & -1 & 0 & 0 & 0 & 0 \\
    0 & 0 & 1 &  1 & 0 & 0 & 0 & 0 \\
    0 & 0 & 0 & 0 & 1 & -1 & 0 & 0 \\
    0 & 0 & 0 & 0 & 1 &  1 & 0 & 0 \\
    0 & 0 & 0 & 0 & 0 & 0 & 1 & -1 \\
    0 & 0 & 0 & 0 & 0 & 0 & 1 &  1 \\
 \end{bmatrix}
\end{equation}
The optimized pulse sequence for this unitary is given in
Table~\ref{taba}. The pulse sequence was obtained with a
theoretical fidelity of 0.995 with a pulse duration
of 107 $\mu$ s. 
The numerically optimized pulse sequence was experimentally
implemented on an initial
thermal equilibrium state, and the result is shown in Fig.~\ref{fig3}.
There is a substantial advantage in the much shorter
duration of the optimized selective pulse which is 
in $\mu$s as compared to the standard pulses which
usually take tens of milliseconds, depending on the
system interactions. The spectra in Figure~\ref{fig3} show
a clean excitation of the third qubit, with no spillover
excitation of the other two qubits.
\subsection{Implementation of CNOT gate}
\label{cnot}
Since the interactions in NMR are always ``on'', it
is most often non-trivial to implement a two-qubit
gate in an $N$ qubit system, while doing nothing
on the other ($N-2$) qubits in the system, as
compared to implementing the same two-qubit
gate in a system of two NMR 
qubits~\cite{linden}.
We hence optimized the two-qubit CNOT gate on our three-qubit
system, with  the first qubit considered the control qubit
while the second qubit was considered the target qubit.  The
corresponding unitary matrix is given by
\begin{equation}
    U_{\rm tgt}=
 \begin{bmatrix}
    1 & 0 & 0 & 0 & 0 & 0 & 0 & 0 \\
    0 & 1 & 0 & 0 & 0 & 0 & 0 & 0 \\
    0 & 0 & 1 & 0 & 0 & 0 & 0 & 0 \\
    0 & 0 & 0 & 1 & 0 & 0 & 0 & 0 \\
    0 & 0 & 0 & 0 & 0 & 0 & 1 & 0 \\
    0 & 0 & 0 & 0 & 0 & 0 & 0 & 1 \\
    0 & 0 & 0 & 0 & 1 & 0 & 0 & 0 \\
    0 & 0 & 0 & 0 & 0 & 1 & 0 & 0 \\
  \end{bmatrix}
\end{equation}
The optimized pulse sequence for this quantum gate is
shown in Table~\ref{tabb},  and was obtained
with a theoretical fidelity of 0.993 with a pulse duration
of 7 ms. 
\begin{figure}[h]
\centering
\includegraphics[angle=0,scale=1.0]{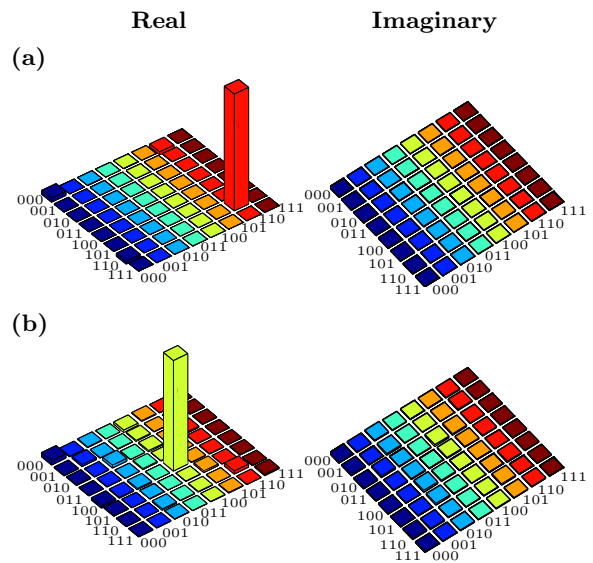}
\caption{Real (left) and imaginary (right) parts of the
experimental tomographs of (a) Initial $\vert110\rangle$
state.
(b) After a CNOT gate applied on the $\vert110\rangle$ state.}
\label{cnottomo}
\end{figure}
\begin{table}[h]
\begin{tabular}[t]{|c|c|c|c|}
\hline
$\mathbf{l}$&$\mathbf{\tau_l(\mu s)}$&$\mathbf{\phi_l}$&$\mathbf{\delta_l(\mu s)}$ \\
\hline   
1& 30 &321.81& 277 \\
2& 33 &320.75& 69\\
3&3&59.02&  1 \\
4&39&66.28&636 \\
5&29&306.06&292 \\
6&9&302.12&19 \\
7&39&312.81&1755\\
8&11&294.12&1 \\
9&39&296.16&636 \\
10&1&157.81&256 \\
\hline   
\end{tabular}
~~~~\begin{tabular}[t]{|c|c|c|c|}
\hline
$\mathbf{l}$&$\mathbf{\tau_l(\mu s)}$&$\mathbf{\phi_l}$&$\mathbf{\delta_l(\mu s)}$ \\
\hline   
11&4&271.97&305 \\
12&39&329.86&11 \\
13&14&352.56&83 \\
14&24&359.89&657 \\
15&1&2.52&1748 \\
16&6&55.02&69 \\
17&4&312.45&2 \\
18&37&309.7&96 \\
\hline
\end{tabular}
\caption{Table representing the pulse sequence for a
two-qubit CNOT
gate. The first column represents the number of propagators.
The second,
third and fourth columns represent the pulse width ($\tau$),
phase ($\phi$) and delay ($\delta$) values, respectively.}
\label{tabb}
\end{table}

The pulse sequence was experimentally implemented on an
initially prepared pseudopure state $\ket{110}$. The final
state was $\ket{100}$ as expected,
with an experimental fidelity of 0.97. The experimentally
tomographed results are shown in Figure~\ref{cnottomo}.

\subsection{Implementation of Fredkin gate}
\label{fredkin}
We optimized the three-qubit Fredkin gate (corresponding to
a controlled-SWAP operation) in a single shot \ie without
breaking it down into other unitaries, and using only a set
of hard pulses and delays. The first qubit was designated as
a control qubit and if the control qubit is 1, then the
other two qubits swap their states. The unitary matrix
corresponding to the Fredkin gate is given by
\begin{equation}
    U_{\rm tgt}=
 \begin{bmatrix}
    1 & 0 & 0 & 0 & 0 & 0 & 0 & 0 \\
    0 & 1 & 0 & 0 & 0 & 0 & 0 & 0 \\
    0 & 0 & 1 & 0 & 0 & 0 & 0 & 0 \\
    0 & 0 & 0 & 1 & 0 & 0 & 0 & 0 \\
    0 & 0 & 0 & 0 & 1 & 0 & 0 & 0 \\
    0 & 0 & 0 & 0 & 0 & 0 & 1 & 0 \\
    0 & 0 & 0 & 0 & 0 & 1 & 0 & 0 \\
    0 & 0 & 0 & 0 & 0 & 0 & 0 & 1 \\
  \end{bmatrix}
\end{equation}
The optimized pulse sequence for this gate is shown in
Table~\ref{tabd}, and
was obtained with a fidelity 0.99 and a pulse
duration of 51 ms. 
\begin{figure}[h]
\centering
\includegraphics[angle=0,scale=1.0]{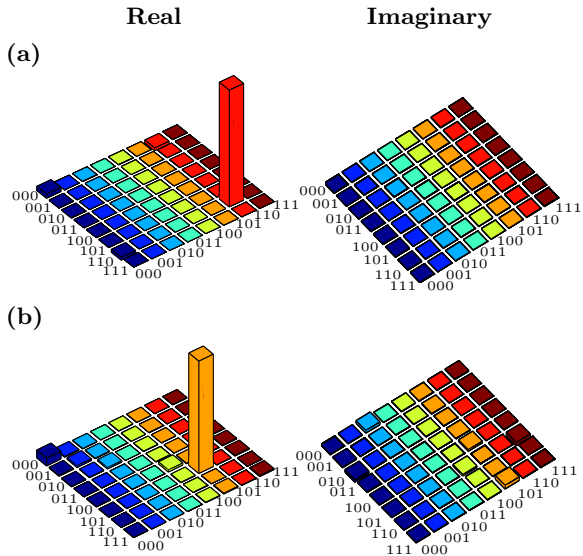}
\caption{Real (left) and imaginary (right) parts of the
experimental tomographs of (a) Initial $\vert110\rangle$
state.
(b) After Fredkin gate
applied on the $\vert110\rangle$ state.}
\label{fredkintomo}
\end{figure}
\begin{table}[h]
\begin{tabular}{|c|c|c|c| }
\hline
$\mathbf{l}$&$\mathbf{\tau_l(\mu s)}$&$\mathbf{\phi_l}$&$\mathbf{\delta_l(ms)}$ \\
 \hline   
1&29&108.63&1.584 \\
2&20&200.75&3.712\\
3&11&258.7&0.639 \\
4&25&197.31&1.075 \\
5&9&172.69&4.568 \\
6&19&188.99&1.964 \\
7&31&265.02&4.416\\
8&27&100.22&2.161 \\
9&12&86.57&3.276 \\
10&32&233.09&2.194\\
\hline
\end{tabular}
~~~~\begin{tabular}{|c|c|c|c| }
\hline
$\mathbf{l}$&$\mathbf{\tau_l(\mu s)}$&$\mathbf{\phi_l}$&$\mathbf{\delta_l(ms)}$ \\
 \hline   
11&10&186&1.569 \\
12&30&170.75&1.592 \\
13&17&4.55&3.269 \\
14&21&188.17&2.184 \\
15&28&36.37&2.152 \\
16&21&330.83&4.423 \\
17&7&46.59&2.194 \\
18&14&102.17&3.066 \\
19&28&295.24&1.574 \\
20&13&126.96&3.720 \\
 \hline
\end{tabular}
\caption{Table representing the pulse sequence for the
Fredkin gate. The first column represents the number of
propagators.  The second, third and fourth columns represent
the pulse width ($\tau$), phase ($\phi$) and delay
($\delta$), respectively.}
\label{tabd}
\end{table}
The pulse sequence was experimentally implemented on
an initial state $\Ket{110}$. The output
state was $\Ket{101}$ with an experimental fidelity of 0.96.
The experimentally tomographed results are shown in 
Figure~\ref{fredkintomo}.

\subsection{Implementation of Toffoli gate}
\label{toffoli}
\begin{figure}[h]
\centering
\includegraphics[angle=0,scale=1.0]{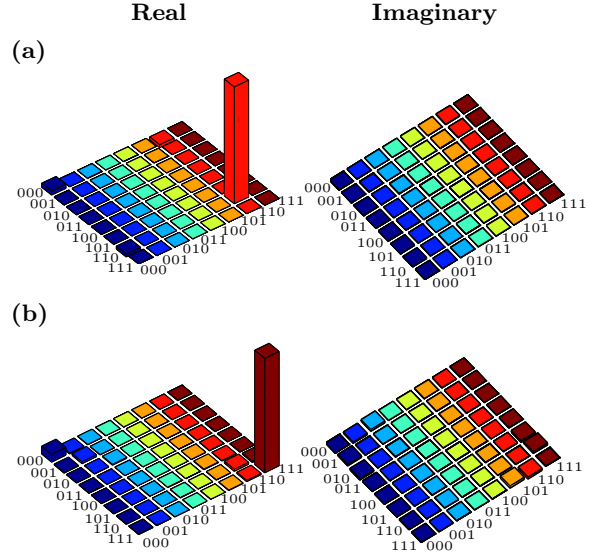}
\caption{Real (left) and imaginary (right) parts of the
experimental tomographs of (a) Initial $\vert110\rangle$
state.
(b)
After a Toffoli gate applied on the $\vert110\rangle$ state.}
\label{toffolitomo}
\end{figure}
The first
two qubits of this gate were considered as the control qubits
while the third qubit was designated the target qubit.  The unitary matrix
corresponding to this gate is given by

\begin{equation}
    U_{\rm tgt}=
  \begin{bmatrix}
    1 & 0 & 0 & 0 & 0 & 0 & 0 & 0 \\
    0 & 1 & 0 & 0 & 0 & 0 & 0 & 0 \\
    0 & 0 & 1 & 0 & 0 & 0 & 0 & 0 \\
    0 & 0 & 0 & 1 & 0 & 0 & 0 & 0 \\
    0 & 0 & 0 & 0 & 1 & 0 & 0 & 0 \\
    0 & 0 & 0 & 0 & 0 & 1 & 0 & 0 \\
    0 & 0 & 0 & 0 & 0 & 0 & 0 & 1 \\
    0 & 0 & 0 & 0 & 0 & 0 & 1 & 0 \\
  \end{bmatrix}
\end{equation}
\begin{figure}[h]
\centering
\includegraphics[angle=0,scale=1.0]{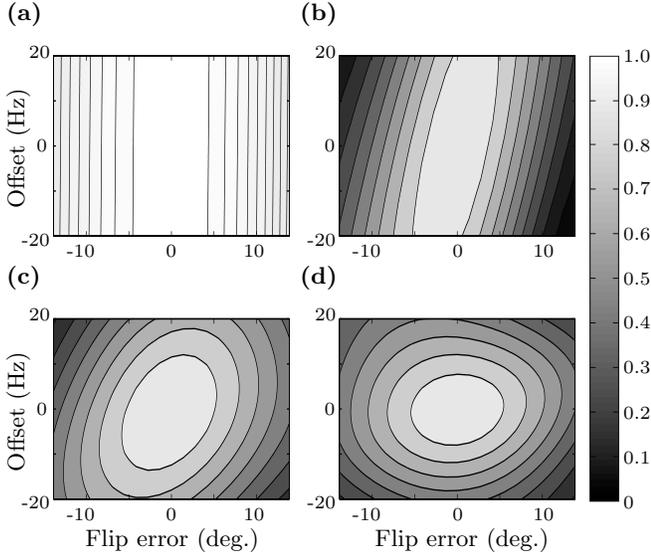}
\caption{Robustness of pulse sequences corresponding to 
(a) $90^{\circ}$ selective pulse  
(b) CNOT gate (c) Fredkin gate and (d) Toffoli gate.
The 
$x$ and $y$ axes represent 
the error in flip angle (deg) and the offset (Hz), respectively. }
\label{fig03}
\end{figure}

\begin{figure}[h]
\centering
\includegraphics[angle=0,scale=1.0]{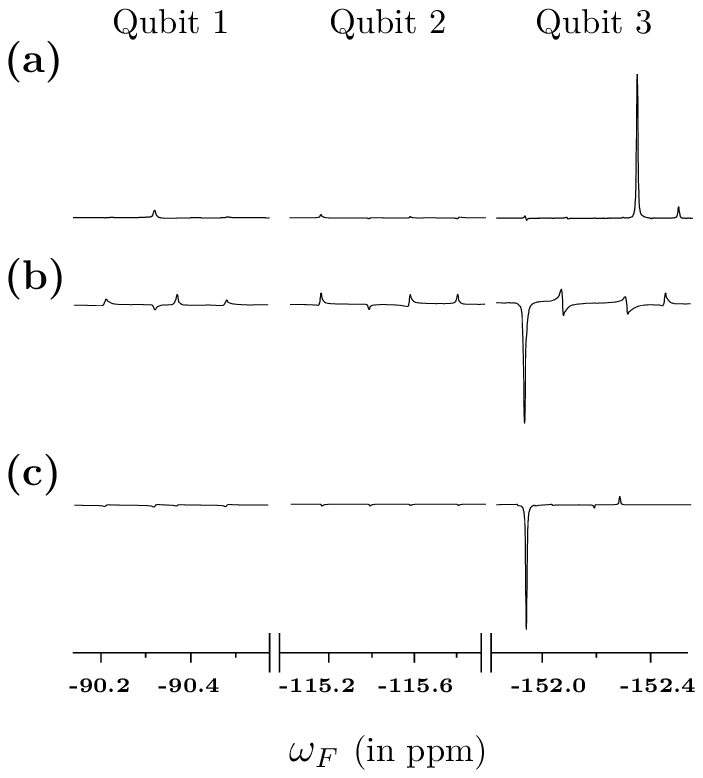}
\caption{
NMR spectra of 
(a) Pseudopure state $\vert 110 \rangle$;  
(b) after the implementation of
a Fredkin gate on the $\vert 110 \rangle$ state 
using transition-selective pulses; (c) after the 
implementation of a Fredkin gate on 
the $\vert 110 \rangle$ state using 
pulses optimized by the genetic algorithm method.
}
\label{stdfredkin}
\end{figure}

The numerically optimized sequence for this gate is shown in 
Table~\ref{tabc} and was obtained with
a fidelity of 0.995 and a pulse duration of 27 ms.
\begin{table}
\begin{tabular}{|c|c|c|c|}
\hline
$\mathbf{l}$&$\mathbf{\tau_l(\mu s)}$&$\mathbf{\phi_l}$&$\mathbf{\delta_l(\mu s)}$ \\
 \hline   
1.&32&243.22&539 \\
2.&27&138.58&546\\
3.&39&2.47&499 \\
4.&36&320.89&3488 \\
5.&32&352.29&2495 \\
6.&34&355.84&536 \\
7.&37&175.98&1938\\
8.&29&20.45&1957 \\
9.&34&354.75&542 \\
10.&18&297.71&564 \\
 \hline
\end{tabular}
~~~\begin{tabular}{|c|c|c|c|}
\hline
$\mathbf{l}$&$\mathbf{\tau_l(\mu s)}$&$\mathbf{\phi_l}$&$\mathbf{\delta_l(\mu s)}$ \\
 \hline   
11.&15&215.55&2487 \\
12.&27&308.2&550 \\
13.&32&326.82&513 \\
14.&13&122.09&541 \\
15.&4&332.61&2518 \\
16.&24&354.12&546 \\
17.&36&310.6&3806 \\
18.&32&210.97&1971 \\
19.&30&3.74&504 \\
20.&38&338.48&565 \\
 \hline
\end{tabular}
\caption{Table representing the pulse sequence for the
Toffoli gate. The first column represents the number of
propagators.
The second, third and fourth columns represent the pulse
width ($\tau$), phase ($\phi$) and delay ($\delta$),
respectively.}
\label{tabc}
\end{table}
The pulse sequence was experimentally implemented on an
initially prepared pseudo-pure state $\Ket{110}$. The final
state was $\Ket{111}$ as expected, and had an experimental
fidelity of 0.93. The experimentally tomographed results are
shown in Figure~\ref{toffolitomo}.

\begin{figure}[h]
\centering
\includegraphics[angle=0,scale=1.0]{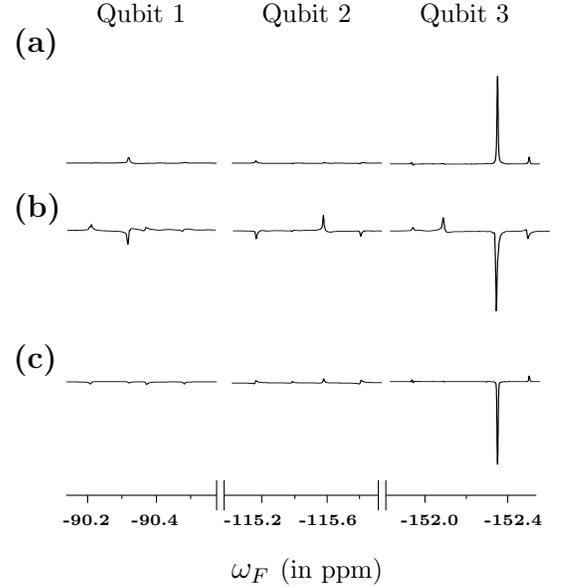}
\caption{
NMR spectra of 
(a) Pseudopure state $\vert 110 \rangle$;  
(b) after the implementation of
a Toffoli gate on the $\vert 110 \rangle$ state 
using transition-selective pulses; (c) after the 
implementation of a Toffoli gate on 
the $\vert 110 \rangle$ state using 
pulses optimized by the genetic algorithm method.
}
\label{stdtoffoli}
\end{figure}

To check the robustness of the numerically optimized pulse sequences we
considered two types of errors: offset errors and flip angle or pulse
miscalibration errors.  Figure~\ref{fig03} shows the variation of fidelity with
the offset frequency (Hz) and flip angle (deg) for the different gates.  We
checked the fidelity variation for the range $\pm 20$ for offset and $\pm 14$
for flip errors.  The $90^{\circ}$ selective rf pulse is most robust of all the
other gates (as is to be expected since it is a single-qubit gate), the
fidelity for this is above 0.90 for the area inside the points  ($\pm12.5$,
$\pm19.6$).  The two-qubit CNOT gate has fidelity $> 0.9$ for the area inside
the points (-5.2,-20), (1.2,-20), (4.7,20) and (-1.2,20). For the Fredkin gate,
the fidelity is $ > 0.9$ for the area which falls under the data points (0,
$\pm 7.8$) and ($\pm 5.3$, 0) which approximately forms an ellipse. The Toffoli
gate has fidelity $> 0.9$ for the area inscribed by points (-2.8, -13.4), (5.5,
0), (1.4,12.2) and (-5.5,0). In general, all the gates optimized by the GA
method are of high fidelity and are robust against both offset and pulse flip
angle errors.

\subsection{Comparison with previous experimental implementations}
\label{compare}
We compared the results of our optimization of the Fredkin
and the Toffoli gates using GAs, with previous experimental
NMR implementations that use transition-selective
pulses~\cite{du-pra-01,fredkin-chinese,dogra-pra-15}.  The
experimental NMR spectra of this comparison are shown in
Figures~\ref{stdfredkin}-\ref{stdtoffoli} for a Fredkin gate
and a Toffoli gate implemented on the $\vert 110 \rangle$
pseudopure state, respectively.  All spectra were recorded
after applying an IIX operation
(\ie a
90$^{\circ}$ pulse on the third qubit).  Since the chemical
shifts of the three fluorine qubits in our particular
molecule cover a very large frequency bandwidth, we crafted
special excitation Gaussian shaped  transition selective
pulses that are frequency modulated~\cite{das-pra-15}.
Using transition-selective pulses, the Fredkin gate was
experimentally implemented with a fidelity of 0.72 and a
pulse length of 242 ms.  Using transition-selective pulses,
the Toffoli gate was experimentally implemented with a
fidelity of 0.76 and a pulse length of 168 ms.  The fidelity
of both the Fredkin and the Toffoli gates using the pulses
crafted using the GA method was $> 0.95$ and the total pulse
durations were substantially smaller, being 51 ms and 27 ms
for the Fredkin and the Toffoli gates, respectively.
Furthermore, as can be seen from the NMR spectra in
Figures~\ref{stdfredkin}-\ref{stdtoffoli}, the standard
implementation of these three-qubit gates using
transition-selective pulses leads to considerable errors due
to decoherence during these long pulses as well as offset
errors. The GA-optimized pulse sequences on the other hand,
have a high fidelity and do not suffer from these errors.
\section{Conclusions}
\label{concl}
In summary, we have optimally designed and experimentally
implemented the universal multi-qubit Toffoli and Fredkin
gates on a three-qubit NMR quantum information processor. We
used a global optimization method based on genetic
algorithms to determine the optimal unitary transformations
and generate the corresponding numerically optimized rf
pulse profiles. We were able to find optimal constructions
for these important three-qubit universal control gates,
which are robust against pulse offset errors as well as
errors that could arise due to decoherence. We were able to
find gate decompositions which are based only on hard (short
duration) rf pulses and delays, which take very short times
to implement and are of high fidelity. Our gate
decompositions are sufficiently general and can be used for
other quantum computing hardwares as well. Although
some of the optimization protocols took a long time to run,
the obvious advantage is that once the optimal pulse sequence for a
gate is found, it can be used later without any further 
optimization
as long as one is working on the same quantum computer.
\acknowledgments
All experiments were performed on Bruker Avance-III 600 MHz
and  400 MHz FT-NMR spectrometers at the NMR Research
Facility at IISER Mohali.  K.D. acknowledges funding from
DST India under Grant number EMR/2015/000556.  A.
acknowledges funding from DST India under Grant number
EMR/2014/000297.  H. S. acknowledges financial support from
CSIR India.  P. P. acknowledges funding from the Indian
Academy of Sciences Bangalore India, under the Summer
Research Fellowship Programme.

\begin{thebibliography}{99}%
\makeatletter
\providecommand \@ifxundefined [1]{%
 \@ifx{#1\undefined}
}%
\providecommand \@ifnum [1]{%
 \ifnum #1\expandafter \@firstoftwo
 \else \expandafter \@secondoftwo
 \fi
}%
\providecommand \@ifx [1]{%
 \ifx #1\expandafter \@firstoftwo
 \else \expandafter \@secondoftwo
 \fi
}%
\providecommand \natexlab [1]{#1}%
\providecommand \enquote  [1]{``#1''}%
\providecommand \bibnamefont  [1]{#1}%
\providecommand \bibfnamefont [1]{#1}%
\providecommand \citenamefont [1]{#1}%
\providecommand \href@noop [0]{\@secondoftwo}%
\providecommand \href [0]{\begingroup \@sanitize@url \@href}%
\providecommand \@href[1]{\@@startlink{#1}\@@href}%
\providecommand \@@href[1]{\endgroup#1\@@endlink}%
\providecommand \@sanitize@url [0]{\catcode `\\12\catcode `\$12\catcode
  `\&12\catcode `\#12\catcode `\^12\catcode `\_12\catcode `\%12\relax}%
\providecommand \@@startlink[1]{}%
\providecommand \@@endlink[0]{}%
\providecommand \url  [0]{\begingroup\@sanitize@url \@url }%
\providecommand \@url [1]{\endgroup\@href {#1}{\urlprefix }}%
\providecommand \urlprefix  [0]{URL }%
\providecommand \Eprint [0]{\href }%
\@ifxundefined \urlstyle {%
  \providecommand \doi  [0]{\begingroup \@sanitize@url \@doi}%
  \providecommand \@doi [1]{\endgroup \@@startlink {\doibase
  #1}doi:\discretionary {}{}{}#1\@@endlink }%
}{%
  \providecommand \doi  [0]{doi:\discretionary{}{}{}\begingroup
  \urlstyle{rm}\Url }%
}%
\providecommand \doibase [0]{http://dx.doi.org/}%
\providecommand \Doi [0]{\begingroup \@sanitize@url \@Doi }%
\providecommand \@Doi  [1]{\endgroup\@@startlink{\doibase#1}\@@Doi}%
\providecommand \@@Doi [1]{#1\@@endlink}%
\providecommand \selectlanguage [0]{\@gobble}%
\providecommand \bibinfo  [0]{\@secondoftwo}%
\providecommand \bibfield  [0]{\@secondoftwo}%
\providecommand \translation [1]{[#1]}%
\providecommand \BibitemOpen [0]{}%
\providecommand \bibitemStop [0]{}%
\providecommand \bibitemNoStop [0]{.\EOS\space}%
\providecommand \EOS [0]{\spacefactor3000\relax}%
\providecommand \BibitemShut  [1]{\csname bibitem#1\endcsname}%
\bibitem [{\citenamefont {Nielsen}\ and\ \citenamefont
  {Chuang}(2000)}]{nielsen-book-02}%
  \BibitemOpen
  \bibfield  {author} {\bibinfo {author} {\bibfnamefont {M.~A.}\ \bibnamefont
  {Nielsen}}\ and\ \bibinfo {author} {\bibfnamefont {I.~L.}\ \bibnamefont
  {Chuang}},\ }\href@noop {} {\emph {\bibinfo {title} {Quantum Computation and
  Quantum Information}}}\ (\bibinfo  {publisher} {Cambridge University Press},\
  \bibinfo {address} {Cambridge UK},\ \bibinfo {year} {2000})\BibitemShut
  {NoStop}%
\bibitem [{\citenamefont {Barenco}\ \emph {et~al.}(1995)\citenamefont
  {Barenco}, \citenamefont {Bennett}, \citenamefont {Cleve}, \citenamefont
  {Divincenzo}, \citenamefont {Margolus}, \citenamefont {Shor}, \citenamefont
  {Sleator}, \citenamefont {Smolin},\ and\ \citenamefont
  {Weinfurter}}]{barenco-pra}%
  \BibitemOpen
  \bibfield  {author} {\bibinfo {author} {\bibfnamefont {A.}~\bibnamefont
  {Barenco}}, \bibinfo {author} {\bibfnamefont {C.~H.}\ \bibnamefont
  {Bennett}}, \bibinfo {author} {\bibfnamefont {R.}~\bibnamefont {Cleve}},
  \bibinfo {author} {\bibfnamefont {D.~P.}\ \bibnamefont {Divincenzo}},
  \bibinfo {author} {\bibfnamefont {N.}~\bibnamefont {Margolus}}, \bibinfo
  {author} {\bibfnamefont {P.}~\bibnamefont {Shor}}, \bibinfo {author}
  {\bibfnamefont {T.}~\bibnamefont {Sleator}}, \bibinfo {author} {\bibfnamefont
  {J.~A.}\ \bibnamefont {Smolin}}, \ and\ \bibinfo {author} {\bibfnamefont
  {H.}~\bibnamefont {Weinfurter}},\ }\href@noop {} {\bibfield  {journal}
  {\bibinfo  {journal} {Phys. Rev. A},\ }\textbf {\bibinfo {volume} {52}},\
  \bibinfo {pages} {3457} (\bibinfo {year} {1995})}\BibitemShut {NoStop}%
\bibitem [{\citenamefont {Sleator}\ and\ \citenamefont
  {Weinfurter}(1995)}]{sleator-prl-95}%
  \BibitemOpen
  \bibfield  {author} {\bibinfo {author} {\bibfnamefont {T.}~\bibnamefont
  {Sleator}}\ and\ \bibinfo {author} {\bibfnamefont {H.}~\bibnamefont
  {Weinfurter}},\ }\href@noop {} {\bibfield  {journal} {\bibinfo  {journal}
  {Phys. Rev. Lett.},\ }\textbf {\bibinfo {volume} {74}},\ \bibinfo {pages}
  {4087} (\bibinfo {year} {1995})}\BibitemShut {NoStop}%
\bibitem [{\citenamefont {Mottonen}\ \emph {et~al.}(2004)\citenamefont
  {Mottonen}, \citenamefont {Vartiainen}, \citenamefont {Bergholm},\ and\
  \citenamefont {Salomaa}}]{mikko-prl-04}%
  \BibitemOpen
  \bibfield  {author} {\bibinfo {author} {\bibfnamefont {M.}~\bibnamefont
  {Mottonen}}, \bibinfo {author} {\bibfnamefont {J.~J.}\ \bibnamefont
  {Vartiainen}}, \bibinfo {author} {\bibfnamefont {V.}~\bibnamefont
  {Bergholm}}, \ and\ \bibinfo {author} {\bibfnamefont {M.~M.}\ \bibnamefont
  {Salomaa}},\ }\href@noop {} {\bibfield  {journal} {\bibinfo  {journal} {Phys.
  Rev. Lett.},\ }\textbf {\bibinfo {volume} {93}},\ \bibinfo {pages} {130502}
  (\bibinfo {year} {2004})}\BibitemShut {NoStop}%
\bibitem [{\citenamefont {Mohammadi}\ and\ \citenamefont
  {Eshghi}(2008)}]{mohammadi-qip-08}%
  \BibitemOpen
  \bibfield  {author} {\bibinfo {author} {\bibfnamefont {M.}~\bibnamefont
  {Mohammadi}}\ and\ \bibinfo {author} {\bibfnamefont {M.}~\bibnamefont
  {Eshghi}},\ }\href@noop {} {\bibfield  {journal} {\bibinfo  {journal} {Quant.
  Inf. Process.},\ }\textbf {\bibinfo {volume} {7}},\ \bibinfo {pages} {175}
  (\bibinfo {year} {2008})}\BibitemShut {NoStop}%
\bibitem [{\citenamefont {Smolin}\ and\ \citenamefont
  {DiVincenzo}(1996)}]{smolin-pra-96}%
  \BibitemOpen
  \bibfield  {author} {\bibinfo {author} {\bibfnamefont {J.~A.}\ \bibnamefont
  {Smolin}}\ and\ \bibinfo {author} {\bibfnamefont {D.~P.}\ \bibnamefont
  {DiVincenzo}},\ }\href@noop {} {\bibfield  {journal} {\bibinfo  {journal}
  {Phys. Rev. A},\ }\textbf {\bibinfo {volume} {53}},\ \bibinfo {pages} {2855}
  (\bibinfo {year} {1996})}\BibitemShut {NoStop}%
\bibitem [{\citenamefont {Buhrman}\ \emph {et~al.}(2001)\citenamefont
  {Buhrman}, \citenamefont {Cleve}, \citenamefont {Watrous},\ and\
  \citenamefont {de~Wolf}}]{buhrman}%
  \BibitemOpen
  \bibfield  {author} {\bibinfo {author} {\bibfnamefont {H.}~\bibnamefont
  {Buhrman}}, \bibinfo {author} {\bibfnamefont {R.}~\bibnamefont {Cleve}},
  \bibinfo {author} {\bibfnamefont {J.}~\bibnamefont {Watrous}}, \ and\
  \bibinfo {author} {\bibfnamefont {R.}~\bibnamefont {de~Wolf}},\ }\href@noop
  {} {\bibfield  {journal} {\bibinfo  {journal} {Phys. Rev. Lett.},\ }\textbf
  {\bibinfo {volume} {87}},\ \bibinfo {pages} {167902} (\bibinfo {year}
  {2001})}\BibitemShut {NoStop}%
\bibitem [{\citenamefont {Hofmann}(2012)}]{hofmann}%
  \BibitemOpen
  \bibfield  {author} {\bibinfo {author} {\bibfnamefont {H.~F.}\ \bibnamefont
  {Hofmann}},\ }\href@noop {} {\bibfield  {journal} {\bibinfo  {journal} {Phys.
  Rev. Lett.},\ }\textbf {\bibinfo {volume} {109}},\ \bibinfo {pages} {020408}
  (\bibinfo {year} {2012})}\BibitemShut {NoStop}%
\bibitem [{\citenamefont {Cory}\ \emph
  {et~al.}(1998){\natexlab{a}}\citenamefont {Cory}, \citenamefont {Price},
  \citenamefont {Maas}, \citenamefont {Knill}, \citenamefont {Laflamme},
  \citenamefont {Zurek}, \citenamefont {Havel},\ and\ \citenamefont
  {Somaroo}}]{cory-qec}%
  \BibitemOpen
  \bibfield  {author} {\bibinfo {author} {\bibfnamefont {D.~G.}\ \bibnamefont
  {Cory}}, \bibinfo {author} {\bibfnamefont {M.~D.}\ \bibnamefont {Price}},
  \bibinfo {author} {\bibfnamefont {W.}~\bibnamefont {Maas}}, \bibinfo {author}
  {\bibfnamefont {E.}~\bibnamefont {Knill}}, \bibinfo {author} {\bibfnamefont
  {R.}~\bibnamefont {Laflamme}}, \bibinfo {author} {\bibfnamefont {W.~H.}\
  \bibnamefont {Zurek}}, \bibinfo {author} {\bibfnamefont {T.~F.}\ \bibnamefont
  {Havel}}, \ and\ \bibinfo {author} {\bibfnamefont {S.~S.}\ \bibnamefont
  {Somaroo}},\ }\href@noop {} {\bibfield  {journal} {\bibinfo  {journal} {Phys.
  Rev. Lett.},\ }\textbf {\bibinfo {volume} {81}},\ \bibinfo {pages} {2152}
  (\bibinfo {year} {1998}{\natexlab{a}})}\BibitemShut {NoStop}%
\bibitem [{\citenamefont {Reed}\ \emph {et~al.}(2012)\citenamefont {Reed},
  \citenamefont {DiCarlo}, \citenamefont {Nigg}, \citenamefont {Sun},
  \citenamefont {Frunzio}, \citenamefont {Girvin},\ and\ \citenamefont
  {Schoelkopf}}]{nature-qec}%
  \BibitemOpen
  \bibfield  {author} {\bibinfo {author} {\bibfnamefont {M.~D.}\ \bibnamefont
  {Reed}}, \bibinfo {author} {\bibfnamefont {L.}~\bibnamefont {DiCarlo}},
  \bibinfo {author} {\bibfnamefont {S.~E.}\ \bibnamefont {Nigg}}, \bibinfo
  {author} {\bibfnamefont {L.}~\bibnamefont {Sun}}, \bibinfo {author}
  {\bibfnamefont {L.}~\bibnamefont {Frunzio}}, \bibinfo {author} {\bibfnamefont
  {S.~M.}\ \bibnamefont {Girvin}}, \ and\ \bibinfo {author} {\bibfnamefont
  {R.~J.}\ \bibnamefont {Schoelkopf}},\ }\href@noop {} {\bibfield  {journal}
  {\bibinfo  {journal} {Nature},\ }\textbf {\bibinfo {volume} {482}},\ \bibinfo
  {pages} {382} (\bibinfo {year} {2012})}\BibitemShut {NoStop}%
\bibitem [{\citenamefont {Milburn}(1989)}]{milburn-prl-89}%
  \BibitemOpen
  \bibfield  {author} {\bibinfo {author} {\bibfnamefont {G.~J.}\ \bibnamefont
  {Milburn}},\ }\href@noop {} {\bibfield  {journal} {\bibinfo  {journal} {Phys.
  Rev. Lett.},\ }\textbf {\bibinfo {volume} {62}},\ \bibinfo {pages} {2124}
  (\bibinfo {year} {1989})}\BibitemShut {NoStop}%
\bibitem [{\citenamefont {Shamir}\ \emph {et~al.}(1986)\citenamefont {Shamir},
  \citenamefont {Caulfield}, \citenamefont {Micelli},\ and\ \citenamefont
  {Seymour}}]{shamir-86}%
  \BibitemOpen
  \bibfield  {author} {\bibinfo {author} {\bibfnamefont {J.}~\bibnamefont
  {Shamir}}, \bibinfo {author} {\bibfnamefont {H.~J.}\ \bibnamefont
  {Caulfield}}, \bibinfo {author} {\bibfnamefont {W.}~\bibnamefont {Micelli}},
  \ and\ \bibinfo {author} {\bibfnamefont {R.~J.}\ \bibnamefont {Seymour}},\
  }\href@noop {} {\bibfield  {journal} {\bibinfo  {journal} {Appl. Opt.},\
  }\textbf {\bibinfo {volume} {25}},\ \bibinfo {pages} {1604} (\bibinfo {year}
  {1986})}\BibitemShut {NoStop}%
\bibitem [{\citenamefont {Peres}(1985)}]{peres-pra-85}%
  \BibitemOpen
  \bibfield  {author} {\bibinfo {author} {\bibfnamefont {A.}~\bibnamefont
  {Peres}},\ }\href@noop {} {\bibfield  {journal} {\bibinfo  {journal} {Phys.
  Rev. A},\ }\textbf {\bibinfo {volume} {32}},\ \bibinfo {pages} {3266}
  (\bibinfo {year} {1985})}\BibitemShut {NoStop}%
\bibitem [{\citenamefont {Shi}(2003)}]{shi-03}%
  \BibitemOpen
  \bibfield  {author} {\bibinfo {author} {\bibfnamefont {Y.}~\bibnamefont
  {Shi}},\ }\href@noop {} {\bibfield  {journal} {\bibinfo  {journal} {Quantum
  Inf. Comput.},\ }\textbf {\bibinfo {volume} {3}},\ \bibinfo {pages} {84}
  (\bibinfo {year} {2003})}\BibitemShut {NoStop}%
\bibitem [{\citenamefont {Chau}\ and\ \citenamefont
  {Wilczek}(1995)}]{chau-prl-95}%
  \BibitemOpen
  \bibfield  {author} {\bibinfo {author} {\bibfnamefont {H.~F.}\ \bibnamefont
  {Chau}}\ and\ \bibinfo {author} {\bibfnamefont {F.}~\bibnamefont {Wilczek}},\
  }\href@noop {} {\bibfield  {journal} {\bibinfo  {journal} {Phys. Rev.
  Lett.},\ }\textbf {\bibinfo {volume} {75}},\ \bibinfo {pages} {748} (\bibinfo
  {year} {1995})}\BibitemShut {NoStop}%
\bibitem [{\citenamefont {Shende}\ and\ \citenamefont
  {Markov}(2009)}]{shende-09}%
  \BibitemOpen
  \bibfield  {author} {\bibinfo {author} {\bibfnamefont {V.~V.}\ \bibnamefont
  {Shende}}\ and\ \bibinfo {author} {\bibfnamefont {I.~L.}\ \bibnamefont
  {Markov}},\ }\href@noop {} {\bibfield  {journal} {\bibinfo  {journal}
  {Quantum Inf. Comput.},\ }\textbf {\bibinfo {volume} {9}},\ \bibinfo {pages}
  {461} (\bibinfo {year} {2009})}\BibitemShut {NoStop}%
\bibitem [{\citenamefont {Ivanov}\ \emph {et~al.}(2015)\citenamefont {Ivanov},
  \citenamefont {Ivanov},\ and\ \citenamefont {Vitanov}}]{ivanov-pra-15}%
  \BibitemOpen
  \bibfield  {author} {\bibinfo {author} {\bibfnamefont {S.~S.}\ \bibnamefont
  {Ivanov}}, \bibinfo {author} {\bibfnamefont {P.~A.}\ \bibnamefont {Ivanov}},
  \ and\ \bibinfo {author} {\bibfnamefont {N.~V.}\ \bibnamefont {Vitanov}},\
  }\Doi {10.1103/PhysRevA.91.032311} {\bibfield  {journal} {\bibinfo  {journal}
  {Phys. Rev. A},\ }\textbf {\bibinfo {volume} {91}},\ \bibinfo {pages}
  {032311} (\bibinfo {year} {2015})}\BibitemShut {NoStop}%
\bibitem [{\citenamefont {Zahedinejad}\ \emph {et~al.}(2015)\citenamefont
  {Zahedinejad}, \citenamefont {Ghosh},\ and\ \citenamefont
  {Sanders}}]{zahedinejad-prl-15}%
  \BibitemOpen
  \bibfield  {author} {\bibinfo {author} {\bibfnamefont {E.}~\bibnamefont
  {Zahedinejad}}, \bibinfo {author} {\bibfnamefont {J.}~\bibnamefont {Ghosh}},
  \ and\ \bibinfo {author} {\bibfnamefont {B.~C.}\ \bibnamefont {Sanders}},\
  }\Doi {10.1103/PhysRevLett.114.200502} {\bibfield  {journal} {\bibinfo
  {journal} {Phys. Rev. Lett.},\ }\textbf {\bibinfo {volume} {114}},\ \bibinfo
  {pages} {200502} (\bibinfo {year} {2015})}\BibitemShut {NoStop}%
\bibitem [{\citenamefont {Zahedinejad}\ \emph {et~al.}(2016)\citenamefont
  {Zahedinejad}, \citenamefont {Ghosh},\ and\ \citenamefont
  {Sanders}}]{sanders-16}%
  \BibitemOpen
  \bibfield  {author} {\bibinfo {author} {\bibfnamefont {E.}~\bibnamefont
  {Zahedinejad}}, \bibinfo {author} {\bibfnamefont {J.}~\bibnamefont {Ghosh}},
  \ and\ \bibinfo {author} {\bibfnamefont {B.~C.}\ \bibnamefont {Sanders}},\
  }\href@noop {} {\bibfield  {journal} {\bibinfo  {journal} {Phys. Rev.
  Appl.},\ }\textbf {\bibinfo {volume} {6}},\ \bibinfo {pages} {054005}
  (\bibinfo {year} {2016})}\BibitemShut {NoStop}%
\bibitem [{\citenamefont {Xue}\ \emph {et~al.}(2002)\citenamefont {Xue},
  \citenamefont {Du}, \citenamefont {Shi}, \citenamefont {Zhou}, \citenamefont
  {Hand},\ and\ \citenamefont {Wu}}]{fredkin-chinese}%
  \BibitemOpen
  \bibfield  {author} {\bibinfo {author} {\bibfnamefont {F.}~\bibnamefont
  {Xue}}, \bibinfo {author} {\bibfnamefont {J.-F.}\ \bibnamefont {Du}},
  \bibinfo {author} {\bibfnamefont {M.-J.}\ \bibnamefont {Shi}}, \bibinfo
  {author} {\bibfnamefont {X.-Y.}\ \bibnamefont {Zhou}}, \bibinfo {author}
  {\bibfnamefont {R.-D.}\ \bibnamefont {Hand}}, \ and\ \bibinfo {author}
  {\bibfnamefont {J.-H.}\ \bibnamefont {Wu}},\ }\href@noop {} {\bibfield
  {journal} {\bibinfo  {journal} {Chin. Phys. Lett.},\ }\textbf {\bibinfo
  {volume} {19}},\ \bibinfo {pages} {1048} (\bibinfo {year}
  {2002})}\BibitemShut {NoStop}%
\bibitem [{\citenamefont {Du}\ \emph {et~al.}(2001)\citenamefont {Du},
  \citenamefont {Shi}, \citenamefont {Wu}, \citenamefont {Zhou},\ and\
  \citenamefont {Han}}]{du-pra-01}%
  \BibitemOpen
  \bibfield  {author} {\bibinfo {author} {\bibfnamefont {J.}~\bibnamefont
  {Du}}, \bibinfo {author} {\bibfnamefont {M.}~\bibnamefont {Shi}}, \bibinfo
  {author} {\bibfnamefont {J.}~\bibnamefont {Wu}}, \bibinfo {author}
  {\bibfnamefont {X.}~\bibnamefont {Zhou}}, \ and\ \bibinfo {author}
  {\bibfnamefont {R.}~\bibnamefont {Han}},\ }\Doi {10.1103/PhysRevA.63.042302}
  {\bibfield  {journal} {\bibinfo  {journal} {Phys. Rev. A},\ }\textbf
  {\bibinfo {volume} {63}},\ \bibinfo {pages} {042302} (\bibinfo {year}
  {2001})}\BibitemShut {NoStop}%
\bibitem [{\citenamefont {Das}\ \emph {et~al.}(2002)\citenamefont {Das},
  \citenamefont {Mahesh},\ and\ \citenamefont {Kumar}}]{das-jmr-02}%
  \BibitemOpen
  \bibfield  {author} {\bibinfo {author} {\bibfnamefont {R.}~\bibnamefont
  {Das}}, \bibinfo {author} {\bibfnamefont {T.}~\bibnamefont {Mahesh}}, \ and\
  \bibinfo {author} {\bibfnamefont {A.}~\bibnamefont {Kumar}},\ }\Doi
  {http://dx.doi.org/10.1016/S1090-7807(02)00009-5} {\bibfield  {journal}
  {\bibinfo  {journal} {J. Magn. Reson.},\ }\textbf {\bibinfo {volume} {159}},\
  \bibinfo {pages} {46 } (\bibinfo {year} {2002})},\ ISSN \bibinfo {issn}
  {1090-7807}\BibitemShut {NoStop}%
\bibitem [{\citenamefont {Zhang}\ \emph {et~al.}(2005)\citenamefont {Zhang},
  \citenamefont {Liu}, \citenamefont {Deng}, \citenamefont {Lu},\ and\
  \citenamefont {Long}}]{zhang-joptb-05}%
  \BibitemOpen
  \bibfield  {author} {\bibinfo {author} {\bibfnamefont {J.}~\bibnamefont
  {Zhang}}, \bibinfo {author} {\bibfnamefont {W.}~\bibnamefont {Liu}}, \bibinfo
  {author} {\bibfnamefont {Z.}~\bibnamefont {Deng}}, \bibinfo {author}
  {\bibfnamefont {Z.}~\bibnamefont {Lu}}, \ and\ \bibinfo {author}
  {\bibfnamefont {G.~L.}\ \bibnamefont {Long}},\ }\href
  {http://stacks.iop.org/1464-4266/7/i=1/a=005} {\bibfield  {journal} {\bibinfo
   {journal} {J. Opt. B},\ }\textbf {\bibinfo {volume} {7}},\ \bibinfo {pages}
  {22} (\bibinfo {year} {2005})}\BibitemShut {NoStop}%
\bibitem [{\citenamefont {Monz}\ \emph {et~al.}(2009)\citenamefont {Monz},
  \citenamefont {Kim}, \citenamefont {Hansel}, \citenamefont {Riebe},
  \citenamefont {Villar}, \citenamefont {Schindler}, \citenamefont {Chwalla},
  \citenamefont {Hennrich},\ and\ \citenamefont {Blatt}}]{monz-prl-09}%
  \BibitemOpen
  \bibfield  {author} {\bibinfo {author} {\bibfnamefont {T.}~\bibnamefont
  {Monz}}, \bibinfo {author} {\bibfnamefont {K.}~\bibnamefont {Kim}}, \bibinfo
  {author} {\bibfnamefont {W.}~\bibnamefont {Hansel}}, \bibinfo {author}
  {\bibfnamefont {M.}~\bibnamefont {Riebe}}, \bibinfo {author} {\bibfnamefont
  {A.~S.}\ \bibnamefont {Villar}}, \bibinfo {author} {\bibfnamefont
  {P.}~\bibnamefont {Schindler}}, \bibinfo {author} {\bibfnamefont
  {M.}~\bibnamefont {Chwalla}}, \bibinfo {author} {\bibfnamefont
  {M.}~\bibnamefont {Hennrich}}, \ and\ \bibinfo {author} {\bibfnamefont
  {R.}~\bibnamefont {Blatt}},\ }\href@noop {} {\bibfield  {journal} {\bibinfo
  {journal} {Phys. Rev. Lett.},\ }\textbf {\bibinfo {volume} {102}},\ \bibinfo
  {pages} {040501} (\bibinfo {year} {2009})}\BibitemShut {NoStop}%
\bibitem [{\citenamefont {Stojanovic}\ \emph {et~al.}(2012)\citenamefont
  {Stojanovic}, \citenamefont {Fedorov}, \citenamefont {Wallraff},\ and\
  \citenamefont {Bruder}}]{qed-prb-12}%
  \BibitemOpen
  \bibfield  {author} {\bibinfo {author} {\bibfnamefont {V.~M.}\ \bibnamefont
  {Stojanovic}}, \bibinfo {author} {\bibfnamefont {A.}~\bibnamefont {Fedorov}},
  \bibinfo {author} {\bibfnamefont {A.}~\bibnamefont {Wallraff}}, \ and\
  \bibinfo {author} {\bibfnamefont {C.}~\bibnamefont {Bruder}},\ }\href@noop {}
  {\bibfield  {journal} {\bibinfo  {journal} {Phys. Rev. B},\ }\textbf
  {\bibinfo {volume} {85}},\ \bibinfo {pages} {054504} (\bibinfo {year}
  {2012})}\BibitemShut {NoStop}%
\bibitem [{\citenamefont {Fedorov}\ \emph {et~al.}(2012)\citenamefont
  {Fedorov}, \citenamefont {Steffen}, \citenamefont {Baur}, \citenamefont
  {da~Silva},\ and\ \citenamefont {Wallraff}}]{toffoli-supercon}%
  \BibitemOpen
  \bibfield  {author} {\bibinfo {author} {\bibfnamefont {A.}~\bibnamefont
  {Fedorov}}, \bibinfo {author} {\bibfnamefont {L.}~\bibnamefont {Steffen}},
  \bibinfo {author} {\bibfnamefont {M.}~\bibnamefont {Baur}}, \bibinfo {author}
  {\bibfnamefont {M.~P.}\ \bibnamefont {da~Silva}}, \ and\ \bibinfo {author}
  {\bibfnamefont {A.}~\bibnamefont {Wallraff}},\ }\href@noop {} {\bibfield
  {journal} {\bibinfo  {journal} {Nature},\ }\textbf {\bibinfo {volume}
  {481}},\ \bibinfo {pages} {170} (\bibinfo {year} {2012})}\BibitemShut
  {NoStop}%
\bibitem [{\citenamefont {Chen}\ \emph {et~al.}(2016)\citenamefont {Chen},
  \citenamefont {Chen},\ and\ \citenamefont {Ma}}]{toffoli-superconducting}%
  \BibitemOpen
  \bibfield  {author} {\bibinfo {author} {\bibfnamefont {M.-F.}\ \bibnamefont
  {Chen}}, \bibinfo {author} {\bibfnamefont {Y.-F.}\ \bibnamefont {Chen}}, \
  and\ \bibinfo {author} {\bibfnamefont {S.-S.}\ \bibnamefont {Ma}},\
  }\href@noop {} {\bibfield  {journal} {\bibinfo  {journal} {Quant. Inf.
  Process.},\ }\textbf {\bibinfo {volume} {15}},\ \bibinfo {pages} {1469}
  (\bibinfo {year} {2016})}\BibitemShut {NoStop}%
\bibitem [{\citenamefont {Lanyon}\ \emph {et~al.}(2009)\citenamefont {Lanyon},
  \citenamefont {Barbieri}, \citenamefont {Almeida}, \citenamefont {Jennewein},
  \citenamefont {Ralph}, \citenamefont {Resch}, \citenamefont {Pryde},
  \citenamefont {O'Brien}, \citenamefont {Gilchrist},\ and\ \citenamefont
  {White}}]{lanyon-nature}%
  \BibitemOpen
  \bibfield  {author} {\bibinfo {author} {\bibfnamefont {B.~P.}\ \bibnamefont
  {Lanyon}}, \bibinfo {author} {\bibfnamefont {M.}~\bibnamefont {Barbieri}},
  \bibinfo {author} {\bibfnamefont {M.~P.}\ \bibnamefont {Almeida}}, \bibinfo
  {author} {\bibfnamefont {T.}~\bibnamefont {Jennewein}}, \bibinfo {author}
  {\bibfnamefont {T.~C.}\ \bibnamefont {Ralph}}, \bibinfo {author}
  {\bibfnamefont {K.~J.}\ \bibnamefont {Resch}}, \bibinfo {author}
  {\bibfnamefont {G.~J.}\ \bibnamefont {Pryde}}, \bibinfo {author}
  {\bibfnamefont {J.~L.}\ \bibnamefont {O'Brien}}, \bibinfo {author}
  {\bibfnamefont {A.}~\bibnamefont {Gilchrist}}, \ and\ \bibinfo {author}
  {\bibfnamefont {A.~G.}\ \bibnamefont {White}},\ }\href@noop {} {\bibfield
  {journal} {\bibinfo  {journal} {Nature Phys.},\ }\textbf {\bibinfo {volume}
  {5}},\ \bibinfo {pages} {134} (\bibinfo {year} {2009})}\BibitemShut {NoStop}%
\bibitem [{\citenamefont {Micuda}\ \emph {et~al.}(2013)\citenamefont {Micuda},
  \citenamefont {Sedlak}, \citenamefont {Straka}, \citenamefont {Mikova},
  \citenamefont {Dusek}, \citenamefont {Jezek},\ and\ \citenamefont
  {Fiurasek}}]{optical-toffoli-prl}%
  \BibitemOpen
  \bibfield  {author} {\bibinfo {author} {\bibfnamefont {M.}~\bibnamefont
  {Micuda}}, \bibinfo {author} {\bibfnamefont {M.}~\bibnamefont {Sedlak}},
  \bibinfo {author} {\bibfnamefont {I.}~\bibnamefont {Straka}}, \bibinfo
  {author} {\bibfnamefont {M.}~\bibnamefont {Mikova}}, \bibinfo {author}
  {\bibfnamefont {M.}~\bibnamefont {Dusek}}, \bibinfo {author} {\bibfnamefont
  {M.}~\bibnamefont {Jezek}}, \ and\ \bibinfo {author} {\bibfnamefont
  {J.}~\bibnamefont {Fiurasek}},\ }\href@noop {} {\bibfield  {journal}
  {\bibinfo  {journal} {Phys. Rev. Lett.},\ }\textbf {\bibinfo {volume}
  {111}},\ \bibinfo {pages} {160407} (\bibinfo {year} {2013})}\BibitemShut
  {NoStop}%
\bibitem [{\citenamefont {Micuda}\ \emph {et~al.}(2015)\citenamefont {Micuda},
  \citenamefont {Mikova}, \citenamefont {Straka}, \citenamefont {Sedlak},
  \citenamefont {Dusek}, \citenamefont {Jezek},\ and\ \citenamefont
  {Fiurasek}}]{optical-toffoli-pra}%
  \BibitemOpen
  \bibfield  {author} {\bibinfo {author} {\bibfnamefont {M.}~\bibnamefont
  {Micuda}}, \bibinfo {author} {\bibfnamefont {M.}~\bibnamefont {Mikova}},
  \bibinfo {author} {\bibfnamefont {I.}~\bibnamefont {Straka}}, \bibinfo
  {author} {\bibfnamefont {M.}~\bibnamefont {Sedlak}}, \bibinfo {author}
  {\bibfnamefont {M.}~\bibnamefont {Dusek}}, \bibinfo {author} {\bibfnamefont
  {M.}~\bibnamefont {Jezek}}, \ and\ \bibinfo {author} {\bibfnamefont
  {J.}~\bibnamefont {Fiurasek}},\ }\href@noop {} {\bibfield  {journal}
  {\bibinfo  {journal} {Phys. Rev. A},\ }\textbf {\bibinfo {volume} {92}},\
  \bibinfo {pages} {032312} (\bibinfo {year} {2015})}\BibitemShut {NoStop}%
\bibitem [{\citenamefont {Moqadam}\ \emph {et~al.}(2016)\citenamefont
  {Moqadam}, \citenamefont {Welter},\ and\ \citenamefont
  {Esquef}}]{optimized-toffoli}%
  \BibitemOpen
  \bibfield  {author} {\bibinfo {author} {\bibfnamefont {J.~K.}\ \bibnamefont
  {Moqadam}}, \bibinfo {author} {\bibfnamefont {G.~S.}\ \bibnamefont {Welter}},
  \ and\ \bibinfo {author} {\bibfnamefont {P.~A.~A.}\ \bibnamefont {Esquef}},\
  }\href@noop {} {\bibfield  {journal} {\bibinfo  {journal} {Quant. Inf.
  Process.},\ }\textbf {\bibinfo {volume} {15}},\ \bibinfo {pages} {4501}
  (\bibinfo {year} {2016})}\BibitemShut {NoStop}%
\bibitem [{\citenamefont {Luo}\ \emph {et~al.}(2015)\citenamefont {Luo},
  \citenamefont {Ma}, \citenamefont {Chen},\ and\ \citenamefont
  {Wang}}]{toffoli-scirep-15}%
  \BibitemOpen
  \bibfield  {author} {\bibinfo {author} {\bibfnamefont {M.-X.}\ \bibnamefont
  {Luo}}, \bibinfo {author} {\bibfnamefont {S.-Y.}\ \bibnamefont {Ma}},
  \bibinfo {author} {\bibfnamefont {X.-B.}\ \bibnamefont {Chen}}, \ and\
  \bibinfo {author} {\bibfnamefont {X.}~\bibnamefont {Wang}},\ }\href@noop {}
  {\bibfield  {journal} {\bibinfo  {journal} {Scientific Reports},\ }\textbf
  {\bibinfo {volume} {5}},\ \bibinfo {pages} {16716} (\bibinfo {year}
  {2015})}\BibitemShut {NoStop}%
\bibitem [{\citenamefont {Ono}\ \emph {et~al.}(2017)\citenamefont {Ono},
  \citenamefont {Okamoto}, \citenamefont {Tanida}, \citenamefont {Hofmann},\
  and\ \citenamefont {Takeuchi}}]{ctrlswap-scirep-17}%
  \BibitemOpen
  \bibfield  {author} {\bibinfo {author} {\bibfnamefont {T.}~\bibnamefont
  {Ono}}, \bibinfo {author} {\bibfnamefont {R.}~\bibnamefont {Okamoto}},
  \bibinfo {author} {\bibfnamefont {M.}~\bibnamefont {Tanida}}, \bibinfo
  {author} {\bibfnamefont {H.~F.}\ \bibnamefont {Hofmann}}, \ and\ \bibinfo
  {author} {\bibfnamefont {S.}~\bibnamefont {Takeuchi}},\ }\href@noop {}
  {\bibfield  {journal} {\bibinfo  {journal} {Scientific Reports},\ }\textbf
  {\bibinfo {volume} {7}},\ \bibinfo {pages} {45353} (\bibinfo {year}
  {2017})}\BibitemShut {NoStop}%
\bibitem [{\citenamefont {Fortunato}\ \emph {et~al.}(2002)\citenamefont
  {Fortunato}, \citenamefont {Pravia}, \citenamefont {Boulant}, \citenamefont
  {Teklemariam}, \citenamefont {Havel},\ and\ \citenamefont
  {Cory}}]{fortunato-jcp-02}%
  \BibitemOpen
  \bibfield  {author} {\bibinfo {author} {\bibfnamefont {E.}~\bibnamefont
  {Fortunato}}, \bibinfo {author} {\bibfnamefont {M.}~\bibnamefont {Pravia}},
  \bibinfo {author} {\bibfnamefont {N.}~\bibnamefont {Boulant}}, \bibinfo
  {author} {\bibfnamefont {G.}~\bibnamefont {Teklemariam}}, \bibinfo {author}
  {\bibfnamefont {T.}~\bibnamefont {Havel}}, \ and\ \bibinfo {author}
  {\bibfnamefont {D.}~\bibnamefont {Cory}},\ }\href@noop {} {\bibfield
  {journal} {\bibinfo  {journal} {J. Chem. Phys.},\ }\textbf {\bibinfo {volume}
  {116}},\ \bibinfo {pages} {7599} (\bibinfo {year} {2002})}\BibitemShut
  {NoStop}%
\bibitem [{\citenamefont {Tosner}\ \emph {et~al.}(2009)\citenamefont {Tosner},
  \citenamefont {Vosegaard}, \citenamefont {Kehlet}, \citenamefont {Khaneja},
  \citenamefont {Glaser},\ and\ \citenamefont {Nielsen}}]{tosner-jmr-09}%
  \BibitemOpen
  \bibfield  {author} {\bibinfo {author} {\bibfnamefont {Z.}~\bibnamefont
  {Tosner}}, \bibinfo {author} {\bibfnamefont {T.}~\bibnamefont {Vosegaard}},
  \bibinfo {author} {\bibfnamefont {C.}~\bibnamefont {Kehlet}}, \bibinfo
  {author} {\bibfnamefont {N.}~\bibnamefont {Khaneja}}, \bibinfo {author}
  {\bibfnamefont {S.~J.}\ \bibnamefont {Glaser}}, \ and\ \bibinfo {author}
  {\bibfnamefont {N.~C.}\ \bibnamefont {Nielsen}},\ }\href@noop {} {\bibfield
  {journal} {\bibinfo  {journal} {J. Magn. Reson.},\ }\textbf {\bibinfo
  {volume} {197}},\ \bibinfo {pages} {120} (\bibinfo {year}
  {2009})}\BibitemShut {NoStop}%
\bibitem [{\citenamefont {Schulte-Herbruggen}\ \emph
  {et~al.}(2012)\citenamefont {Schulte-Herbruggen}, \citenamefont {Marx},
  \citenamefont {Fahmy}, \citenamefont {Kauffman}, \citenamefont {Lomonaco},
  \citenamefont {Khaneja},\ and\ \citenamefont {Glaser}}]{glaser-review}%
  \BibitemOpen
  \bibfield  {author} {\bibinfo {author} {\bibfnamefont {T.}~\bibnamefont
  {Schulte-Herbruggen}}, \bibinfo {author} {\bibfnamefont {R.}~\bibnamefont
  {Marx}}, \bibinfo {author} {\bibfnamefont {A.}~\bibnamefont {Fahmy}},
  \bibinfo {author} {\bibfnamefont {L.}~\bibnamefont {Kauffman}}, \bibinfo
  {author} {\bibfnamefont {S.}~\bibnamefont {Lomonaco}}, \bibinfo {author}
  {\bibfnamefont {N.}~\bibnamefont {Khaneja}}, \ and\ \bibinfo {author}
  {\bibfnamefont {S.~J.}\ \bibnamefont {Glaser}},\ }\href@noop {} {\bibfield
  {journal} {\bibinfo  {journal} {Phil. T. Roy. Soc. A},\ }\textbf {\bibinfo
  {volume} {370}},\ \bibinfo {pages} {4651} (\bibinfo {year}
  {2012})}\BibitemShut {NoStop}%
\bibitem [{\citenamefont {Kosut}\ \emph {et~al.}(2013)\citenamefont {Kosut},
  \citenamefont {Grace},\ and\ \citenamefont {Brif}}]{qtmgates-13}%
  \BibitemOpen
  \bibfield  {author} {\bibinfo {author} {\bibfnamefont {R.~L.}\ \bibnamefont
  {Kosut}}, \bibinfo {author} {\bibfnamefont {M.~D.}\ \bibnamefont {Grace}}, \
  and\ \bibinfo {author} {\bibfnamefont {C.}~\bibnamefont {Brif}},\ }\href@noop
  {} {\bibfield  {journal} {\bibinfo  {journal} {Phys. Rev. A},\ }\textbf
  {\bibinfo {volume} {88}},\ \bibinfo {pages} {052326} (\bibinfo {year}
  {2013})}\BibitemShut {NoStop}%
\bibitem [{\citenamefont {Zhang}\ \emph {et~al.}(2015)\citenamefont {Zhang},
  \citenamefont {Wu}, \citenamefont {Zhang}, \citenamefont {Tarn},\ and\
  \citenamefont {Long}}]{long-homonuclear}%
  \BibitemOpen
  \bibfield  {author} {\bibinfo {author} {\bibfnamefont {T.-M.}\ \bibnamefont
  {Zhang}}, \bibinfo {author} {\bibfnamefont {R.-B.}\ \bibnamefont {Wu}},
  \bibinfo {author} {\bibfnamefont {F.-H.}\ \bibnamefont {Zhang}}, \bibinfo
  {author} {\bibfnamefont {T.-J.}\ \bibnamefont {Tarn}}, \ and\ \bibinfo
  {author} {\bibfnamefont {G.-L.}\ \bibnamefont {Long}},\ }\href@noop {}
  {\bibfield  {journal} {\bibinfo  {journal} {IEEE Trans. Ctrl. Syst. Tech.},\
  }\textbf {\bibinfo {volume} {23}},\ \bibinfo {pages} {2018} (\bibinfo {year}
  {2015})}\BibitemShut {NoStop}%
\bibitem [{\citenamefont {Holland}(1992)}]{holland-book}%
  \BibitemOpen
  \bibfield  {author} {\bibinfo {author} {\bibfnamefont {J.~H.}\ \bibnamefont
  {Holland}},\ }\href@noop {} {\emph {\bibinfo {title} {Adaptation in Natural
  and Artificial Systems: An Introductory Analysis with Applications to
  Biology, Control, and Artificial Intelligence}}}\ (\bibinfo  {publisher} {MIT
  Press},\ \bibinfo {address} {Boston USA},\ \bibinfo {year}
  {1992})\BibitemShut {NoStop}%
\bibitem [{\citenamefont {Forrest}(1993)}]{forrest}%
  \BibitemOpen
  \bibfield  {author} {\bibinfo {author} {\bibfnamefont {S.}~\bibnamefont
  {Forrest}},\ }\href@noop {} {\bibfield  {journal} {\bibinfo  {journal}
  {Science},\ }\textbf {\bibinfo {volume} {261}},\ \bibinfo {pages} {872}
  (\bibinfo {year} {1993})}\BibitemShut {NoStop}%
\bibitem [{\citenamefont {Stadelhofer}\ \emph {et~al.}(2008)\citenamefont
  {Stadelhofer}, \citenamefont {Banzhaf},\ and\ \citenamefont
  {Suter}}]{suter-ia-08}%
  \BibitemOpen
  \bibfield  {author} {\bibinfo {author} {\bibfnamefont {R.}~\bibnamefont
  {Stadelhofer}}, \bibinfo {author} {\bibfnamefont {W.}~\bibnamefont
  {Banzhaf}}, \ and\ \bibinfo {author} {\bibfnamefont {D.}~\bibnamefont
  {Suter}},\ }\href@noop {} {\bibfield  {journal} {\bibinfo  {journal} {Art.
  Int. Engg. Des. Anal. Manufac.},\ }\textbf {\bibinfo {volume} {22}},\
  \bibinfo {pages} {285} (\bibinfo {year} {2008})}\BibitemShut {NoStop}%
\bibitem [{\citenamefont {Hardy}\ and\ \citenamefont
  {Steeb}(2010)}]{hardy-2010}%
  \BibitemOpen
  \bibfield  {author} {\bibinfo {author} {\bibfnamefont {Y.}~\bibnamefont
  {Hardy}}\ and\ \bibinfo {author} {\bibfnamefont {W.-H.}\ \bibnamefont
  {Steeb}},\ }\href@noop {} {\bibfield  {journal} {\bibinfo  {journal} {Intl.
  J. Mod. Phys. C},\ }\textbf {\bibinfo {volume} {21}},\ \bibinfo {pages}
  {1359} (\bibinfo {year} {2010})}\BibitemShut {NoStop}%
\bibitem [{\citenamefont {Bang}\ and\ \citenamefont {Yoo}(2014)}]{bang-korean}%
  \BibitemOpen
  \bibfield  {author} {\bibinfo {author} {\bibfnamefont {J.}~\bibnamefont
  {Bang}}\ and\ \bibinfo {author} {\bibfnamefont {S.}~\bibnamefont {Yoo}},\
  }\href@noop {} {\bibfield  {journal} {\bibinfo  {journal} {J. Korean Phys.
  Soc.},\ }\textbf {\bibinfo {volume} {65}},\ \bibinfo {pages} {2001} (\bibinfo
  {year} {2014})}\BibitemShut {NoStop}%
\bibitem [{\citenamefont {Navarro-Munoz}\ \emph {et~al.}(2006)\citenamefont
  {Navarro-Munoz}, \citenamefont {Rosu},\ and\ \citenamefont
  {Lopez-Sandoval}}]{navarro-pra-06}%
  \BibitemOpen
  \bibfield  {author} {\bibinfo {author} {\bibfnamefont {J.~C.}\ \bibnamefont
  {Navarro-Munoz}}, \bibinfo {author} {\bibfnamefont {H.~C.}\ \bibnamefont
  {Rosu}}, \ and\ \bibinfo {author} {\bibfnamefont {R.}~\bibnamefont
  {Lopez-Sandoval}},\ }\href@noop {} {\bibfield  {journal} {\bibinfo  {journal}
  {Phys. Rev. A},\ }\textbf {\bibinfo {volume} {74}},\ \bibinfo {pages}
  {052308} (\bibinfo {year} {2006})}\BibitemShut {NoStop}%
\bibitem [{\citenamefont {Quiroz}\ and\ \citenamefont
  {Lidar}(2013)}]{lidar-pra}%
  \BibitemOpen
  \bibfield  {author} {\bibinfo {author} {\bibfnamefont {G.}~\bibnamefont
  {Quiroz}}\ and\ \bibinfo {author} {\bibfnamefont {D.~A.}\ \bibnamefont
  {Lidar}},\ }\href@noop {} {\bibfield  {journal} {\bibinfo  {journal} {Phys.
  Rev. A},\ }\textbf {\bibinfo {volume} {88}},\ \bibinfo {pages} {052306}
  (\bibinfo {year} {2013})}\BibitemShut {NoStop}%
\bibitem [{\citenamefont {Manu}\ and\ \citenamefont
  {Kumar}(2012)}]{manu-pra-12}%
  \BibitemOpen
  \bibfield  {author} {\bibinfo {author} {\bibfnamefont {V.~S.}\ \bibnamefont
  {Manu}}\ and\ \bibinfo {author} {\bibfnamefont {A.}~\bibnamefont {Kumar}},\
  }\Doi {10.1103/PhysRevA.86.022324} {\bibfield  {journal} {\bibinfo  {journal}
  {Phys. Rev. A},\ }\textbf {\bibinfo {volume} {86}},\ \bibinfo {pages}
  {022324} (\bibinfo {year} {2012})}\BibitemShut {NoStop}%
\bibitem [{\citenamefont {Manu}\ and\ \citenamefont
  {Kumar}(2014)}]{manu-pra-14}%
  \BibitemOpen
  \bibfield  {author} {\bibinfo {author} {\bibfnamefont {V.~S.}\ \bibnamefont
  {Manu}}\ and\ \bibinfo {author} {\bibfnamefont {A.}~\bibnamefont {Kumar}},\
  }\href@noop {} {\bibfield  {journal} {\bibinfo  {journal} {Phys. Rev. A},\
  }\textbf {\bibinfo {volume} {89}},\ \bibinfo {pages} {052331} (\bibinfo
  {year} {2014})}\BibitemShut {NoStop}%
\bibitem [{\citenamefont {Souza}\ \emph {et~al.}(2011)\citenamefont {Souza},
  \citenamefont {Alvarez},\ and\ \citenamefont {Suter}}]{souza-prl-11}%
  \BibitemOpen
  \bibfield  {author} {\bibinfo {author} {\bibfnamefont {A.~M.}\ \bibnamefont
  {Souza}}, \bibinfo {author} {\bibfnamefont {G.~A.}\ \bibnamefont {Alvarez}},
  \ and\ \bibinfo {author} {\bibfnamefont {D.}~\bibnamefont {Suter}},\
  }\href@noop {} {\bibfield  {journal} {\bibinfo  {journal} {Phys. Rev.
  Lett.},\ }\textbf {\bibinfo {volume} {106}},\ \bibinfo {pages} {240501}
  (\bibinfo {year} {2011})}\BibitemShut {NoStop}%
\bibitem [{\citenamefont {Jebari}\ and\ \citenamefont
  {Madiafi}(2013)}]{jebari-selection}%
  \BibitemOpen
  \bibfield  {author} {\bibinfo {author} {\bibfnamefont {K.}~\bibnamefont
  {Jebari}}\ and\ \bibinfo {author} {\bibfnamefont {M.}~\bibnamefont
  {Madiafi}},\ }\href@noop {} {\bibfield  {journal} {\bibinfo  {journal} {Int.
  J. Emerg. Sci.},\ }\textbf {\bibinfo {volume} {3}},\ \bibinfo {pages} {333}
  (\bibinfo {year} {2013})}\BibitemShut {NoStop}%
\bibitem [{\citenamefont {MATLAB}(2015)}]{matlab}%
  \BibitemOpen
  \bibfield  {author} {\bibinfo {author} {\bibnamefont {MATLAB}},\ }\href@noop
  {} {\emph {\bibinfo {title} {Version 8.5.0 (R2015a)}}}\ (\bibinfo
  {publisher} {MathWorks Inc.},\ \bibinfo {address} {Natick, Massachusetts},\
  \bibinfo {year} {2015})\BibitemShut {NoStop}%
\bibitem [{\citenamefont {Cory}\ \emph
  {et~al.}(1998){\natexlab{b}}\citenamefont {Cory}, \citenamefont {Price},\
  and\ \citenamefont {Havel}}]{cory-physicad}%
  \BibitemOpen
  \bibfield  {author} {\bibinfo {author} {\bibfnamefont {D.~G.}\ \bibnamefont
  {Cory}}, \bibinfo {author} {\bibfnamefont {M.~D.}\ \bibnamefont {Price}}, \
  and\ \bibinfo {author} {\bibfnamefont {T.~F.}\ \bibnamefont {Havel}},\
  }\href@noop {} {\bibfield  {journal} {\bibinfo  {journal} {Physica D},\
  }\textbf {\bibinfo {volume} {120}},\ \bibinfo {pages} {82} (\bibinfo {year}
  {1998}{\natexlab{b}})}\BibitemShut {NoStop}%
\bibitem [{\citenamefont {Leskowitz}\ and\ \citenamefont
  {Mueller}(2004)}]{leskowitz}%
  \BibitemOpen
  \bibfield  {author} {\bibinfo {author} {\bibfnamefont {G.~M.}\ \bibnamefont
  {Leskowitz}}\ and\ \bibinfo {author} {\bibfnamefont {L.~J.}\ \bibnamefont
  {Mueller}},\ }\Doi {10.1103/PhysRevA.69.052302} {\bibfield  {journal}
  {\bibinfo  {journal} {Phys. Rev. A},\ }\textbf {\bibinfo {volume} {69}},\
  \bibinfo {pages} {052302} (\bibinfo {year} {2004})}\BibitemShut {NoStop}%
\bibitem [{\citenamefont {Singh}\ \emph {et~al.}(2016)\citenamefont {Singh},
  \citenamefont {Arvind},\ and\ \citenamefont {Dorai}}]{singh-pla-16}%
  \BibitemOpen
  \bibfield  {author} {\bibinfo {author} {\bibfnamefont {H.}~\bibnamefont
  {Singh}}, \bibinfo {author} {\bibnamefont {Arvind}}, \ and\ \bibinfo {author}
  {\bibfnamefont {K.}~\bibnamefont {Dorai}},\ }\Doi
  {http://dx.doi.org/10.1016/j.physleta.2016.07.046} {\bibfield  {journal}
  {\bibinfo  {journal} {Phys. Lett. A},\ }\textbf {\bibinfo {volume} {380}},\
  \bibinfo {pages} {3051 } (\bibinfo {year} {2016})},\ ISSN \bibinfo {issn}
  {0375-9601}\BibitemShut {NoStop}%
\bibitem [{\citenamefont {Uhlmann}(1976)}]{uhlmann}%
  \BibitemOpen
  \bibfield  {author} {\bibinfo {author} {\bibfnamefont {A.}~\bibnamefont
  {Uhlmann}},\ }\Doi {https://doi.org/10.1016/0034-4877(76)90060-4} {\bibfield
  {journal} {\bibinfo  {journal} {Rep. Math. Phys.},\ }\textbf {\bibinfo
  {volume} {9}},\ \bibinfo {pages} {273} (\bibinfo {year} {1976})}\BibitemShut
  {NoStop}%
\bibitem [{\citenamefont {Jozsa}(1994)}]{jozsa}%
  \BibitemOpen
  \bibfield  {author} {\bibinfo {author} {\bibfnamefont {R.}~\bibnamefont
  {Jozsa}},\ }\Doi {10.1080/09500349414552171} {\bibfield  {journal} {\bibinfo
  {journal} {J. Mod. Optics},\ }\textbf {\bibinfo {volume} {41}},\ \bibinfo
  {pages} {2315} (\bibinfo {year} {1994})}\BibitemShut {NoStop}%
\bibitem [{\citenamefont {Linden}\ \emph {et~al.}(1999)\citenamefont {Linden},
  \citenamefont {Herve}, \citenamefont {Carbajo},\ and\ \citenamefont
  {Freeman}}]{linden}%
  \BibitemOpen
  \bibfield  {author} {\bibinfo {author} {\bibfnamefont {N.}~\bibnamefont
  {Linden}}, \bibinfo {author} {\bibfnamefont {B.}~\bibnamefont {Herve}},
  \bibinfo {author} {\bibfnamefont {R.~J.}\ \bibnamefont {Carbajo}}, \ and\
  \bibinfo {author} {\bibfnamefont {R.}~\bibnamefont {Freeman}},\ }\href@noop
  {} {\bibfield  {journal} {\bibinfo  {journal} {Chem. Phys. Lett.},\ }\textbf
  {\bibinfo {volume} {305}},\ \bibinfo {pages} {28} (\bibinfo {year}
  {1999})}\BibitemShut {NoStop}%
\bibitem [{\citenamefont {Dogra}\ \emph {et~al.}(2015)\citenamefont {Dogra},
  \citenamefont {Dorai},\ and\ \citenamefont {Arvind}}]{dogra-pra-15}%
  \BibitemOpen
  \bibfield  {author} {\bibinfo {author} {\bibfnamefont {S.}~\bibnamefont
  {Dogra}}, \bibinfo {author} {\bibfnamefont {K.}~\bibnamefont {Dorai}}, \ and\
  \bibinfo {author} {\bibnamefont {Arvind}},\ }\Doi
  {10.1103/PhysRevA.91.022312} {\bibfield  {journal} {\bibinfo  {journal}
  {Phys. Rev. A},\ }\textbf {\bibinfo {volume} {91}},\ \bibinfo {pages}
  {022312} (\bibinfo {year} {2015})}\BibitemShut {NoStop}%
\bibitem [{\citenamefont {Das}\ \emph {et~al.}(2015)\citenamefont {Das},
  \citenamefont {Dogra}, \citenamefont {Dorai},\ and\ \citenamefont
  {Arvind}}]{das-pra-15}%
  \BibitemOpen
  \bibfield  {author} {\bibinfo {author} {\bibfnamefont {D.}~\bibnamefont
  {Das}}, \bibinfo {author} {\bibfnamefont {S.}~\bibnamefont {Dogra}}, \bibinfo
  {author} {\bibfnamefont {K.}~\bibnamefont {Dorai}}, \ and\ \bibinfo {author}
  {\bibnamefont {Arvind}},\ }\Doi {10.1103/PhysRevA.92.022307} {\bibfield
  {journal} {\bibinfo  {journal} {Phys. Rev. A},\ }\textbf {\bibinfo {volume}
  {92}},\ \bibinfo {pages} {022307} (\bibinfo {year} {2015})}\BibitemShut
  {NoStop}%
\end{thebibliography}
%
\end{document}